\documentclass[12pt]{article}

\usepackage{amsmath,amsfonts,epsfig,color,latexsym,graphics}

\newcommand{\be}{\begin{equation}}
\newcommand{\ee}{\end{equation}}
\newcommand{\bea}{\begin{eqnarray}}
\newcommand{\eea}{\end{eqnarray}}

\newcommand{\CR}{\nonumber\\}

\let\newsection=\section
\renewcommand{\section}{\setcounter{equation}{0}\newsection}
\def\e{\epsilon}
\def\m{\mu}

\begin{document}

\begin{flushright}
TIT/HEP-575\\
October 2007
\end{flushright}
\vskip.5in

\begin{center}

{\LARGE\bf Comments on gluon 6-point scattering amplitudes in N=4 SYM at 
strong coupling }
\vskip 1in
\centerline{\Large Dumitru Astefanesei$^1$, Suguru Dobashi$^1$, Katsushi Ito$^2$ 
and Horatiu Nastase$^1$}
\vskip .5in

\end{center}

\centerline{\large $^1$Global Edge Institute, Tokyo Institute of Technology}
\centerline{\large Ookayama 2-12-1, Meguro, Tokyo 152-8550, Japan}
\vskip .5in

\centerline{\large $^2$Department of Physics, Tokyo Institute of Technology}
\centerline{\large Ookayama 2-12-1, Meguro, Tokyo 152-8551, Japan}

\vskip 1in

\begin{abstract}
{\large
We use the AdS/CFT prescription of Alday and Maldacena \cite{am} to analyze gluon 
6-point scattering amplitudes at strong coupling in ${\cal N}=4$ SYM.  By cutting and 
gluing we obtain AdS 6-point amplitudes that contain extra boundary conditions and come 
close to matching the field theory results. We interpret them as parts of the field theory
amplitudes, containing only certain diagrams. We also analyze the collinear limits of 
6- and 5-point amplitudes and discuss the results.}

\end{abstract}

\newpage

\section{Introduction}

The gauge/gravity duality is a valuable tool to investigate the dynamics of gauge theories. 
Many nonperturbative aspects of gauge theory have been elucidated, mostly for the 
supersymmetric cases, like ${\cal N}=4$ SYM, where many correlations functions have been
analyzed. The thermodynamic and qualitative properties of a large class of gauge theories have been 
obtained. But until recently, only the properties of gauge invariant states were obtained 
this way. In \cite{am} however, the amplitudes for scattering of gluons (coloured states)
in ${\cal N}=4$ SYM were described using AdS/CFT.

${\cal N}=4$ SYM and QCD have quite different dynamics at large distances but there are similarities at short distances.
The perturbative SYM scattering amplitudes have many features in common  with their QCD 
counterparts, e.g. 
the SYM loop amplitudes can be considered as components of QCD loop amplitudes (see \cite{Bern:2004kq} and references therein). It is thus important to learn 
as much as possible about the amplitudes of ${\cal N}=4$ SYM, and hope that we can
extract information that will be relevant for understanding the QCD
physics at hadron colliders.

Alday and Maldacena \cite{am} proposed a method for computing gluon scattering amplitudes at strong coupling in ${\cal N}=4$ SYM. 
The essential feature that allowed for the calculation 
of this coloured amplitude is the factorization of all colour indices into the
tree amplitude, ${\cal A}={\cal A}_{tree}{\cal M}$, the scalar function ${\cal M}$ being 
calculated from the areas of worldsheets of a classical string in a T dual AdS 
space. Classical strings are familiar in AdS/CFT from the calculation of Wilson loops.
Also, large semiclassical strings correspond to gauge theory operators with large
angular momentum \cite{Gubser:2002tv}, or large R-charge and spin chain momentum \cite{hm},
whereas quantum strings correspond to large R charge \cite{bmn}.\footnote{The conventional AdS/CFT correspondence relates the strong coupling regime of ${\cal N}=4$ SYM to the supergravity limit of 
string theory on the $AdS_5 \times S^5$ background for small operators. The analysis of string theory requires large gauge theory operators or, in the spirit of the original 't Hooft string worldsheet 
proposal, analyzing the zero coupling limit \cite{gopa,nas}.}

Now, the string worldsheet has boundary conditions defined by the gluon states.
Gluon states are open strings that end on an infrared D3-brane.
The `T duality on AdS space'
was used as a mathematical trick, mapping the open string worldsheet with
vertex operators defined by external gauge theory momenta to an open string worldsheet 
with usual Dirichlet boundary conditions, defined by lightlike segments forming a 
closed contour, but the AdS space is still noncompact.
After the T duality, the boundary and the infrared region are 
interchanged and so the brane is located on the boundary in the T dual AdS, giving 
formally the same calculation as for a lighlike Wilson loop.  
The 
D3-brane is an infrared regulator in the gauge theory,
needed since gluon amplitudes are IR divergent. 

Using this prescription, \cite{am} computed the 4-point gluon scattering amplitude at strong 
coupling in  ${\cal N}=4$ at large $N$, and 
compared it with the conjectured exact result of Bern, Dixon, and Smirnov (BDS) \cite{bds}
(see also \cite{abdk}). The 
BDS conjecture 
states that the planar contributions to scattering amplitudes of ${\cal N}=4$ SYM have an 
iterative structure (at least, for MHV amplitudes): the 
higher-loop amplitudes are determined by the one-loop amplitude and some 
functions of the coupling constant. 

An $n$-point amplitude factorizes in two parts: an universal 
function  depending  just on momentum 
invariants times the tree-level amplitude that contains all colors and helicity 
factors. Unlike in QCD where the scattering amplitudes are very complicated objects, in a SUSY 
theory the kinematic part is a simple exponential. The four gluon amplitude in ${\cal 
N}=4$ SYM contains an infrared divergent part plus a finite part that is an elementary function 
(log squared) of Mandelstam kinematic variables, 
and is determined by only 
two functions of the 't Hooft coupling. Thus the only nontrivial information is 
encoded in these two functions, one of which is related to the 
cusp anomalous dimension.\footnote{A nice physical interpretation 
of the cusp anomaly at weak coupling within the radial quantization approach was given in 
\cite{bgk}. That is a {\it quantum transition amplitude} for a test particle propagating in the 
radial time and the angular coordinates. Thus, this is an important hint that at strong 
coupling the correspondent quantity is the {\it classical action} for a particle propagating on 
the same phase space.} 

The four gluon scattering amplitude computed at strong coupling has the 
same form as the BDS amplitude at weak coupling with the cusp anomalous value obtained from the 
semiclassical analysis of \cite{Gubser:2002tv}(see also \cite{am2}). Even if the factorization 
does not hold order by order in the coupling constant for non-MHV amplitudes, it holds again in 
the strong coupling limit \cite{am, afk}. 

One possible reason for the simple form
of the conjectured BDS result was explored in \cite{dks}: hidden conformal symmetry of 
the amplitude, not related in an obvious way with the conformal symmetry of the ${\cal 
N}=4$ SYM. Motivated by the work of Alday and Maldacena the authors of \cite{dks,bht} investigated the lightlike Wilson loop at weak coupling. They concluded that the 
duality between gluon amplitudes and Wilson loops is also valid at weak coupling.
This is possible evidence for the hidden conformal symmetry of the ${\cal N}=4$ SYM.
Other recent papers discussing aspects of the Alday-Maldacena proposal are 
\cite{ns,mmt,buch,others}.

In this paper we extend the work of \cite{am} by analyzing  6-point amplitudes. 
It was explained in \cite{dhks,am3} that 4- and 5-point amplitudes are fixed by conformal 
symmetry, and therefore any real test of the BDS conjecture will come for $n=6$ point 
amplitudes and higher
(a conformal Ward identity found in \cite{dhks} fixes the form of the 
4- and 5-point amplitudes, but not higher). In fact, \cite{am3} found that a large $n$ calculation gives 
dissagreement. It is therefore very important to analyze 6-point amplitudes.

We first
calculate the strong coupling prediction of the 6-point amplitudes using the BDS conjecture.
We then construct 6-point AdS amplitudes by using symmetries and
cutting and gluing the 4-point solution. We will see that the lines where we cut and glue 
actually contain extra boundary conditions, and we will try to interpret them in 
gauge theory. We will find an interesting relation of these amplitudes
to the unitarity cut procedure. The gauge theory 6-point 
amplitudes we are studying do not have the most general 
external momenta, and in fact we will obtain a Regge-like behaviour for amplitudes when
some of the momentum invariants go to infinity, while others are fixed, similar to the 
4-point function behaviour checked by \cite{ns}. 

We will also treat for completeness an 8-point  AdS amplitude that can be 
obtained by the same methods, and interpret it in gauge theory. Finally, we will 
look at the collinear behaviour of the 6- and 5-point amplitudes to go to the 4-point 
amplitude. The prescription of \cite{am} implies that it should be possible to get a 
smooth limit, and we comment how that could be achieved.

The paper is organized as follows: In section 2 we review the calculation of \cite{am}.
In section 3 we calculate 6-point amplitudes: first we specify the field theory results, 
and then we calculate the AdS result and compare. In section 4 we interpret the mismatch
and give a gauge theory interpretation of the result. In section 5 we calculate the 
8-point amplitude and  in section 6 we analyze the collinear limit. An appendix gives some
calculational details.

\section{Review}

Alday and Maldacena \cite{am}
describe the 4 dimensional 2 to 2 scattering amplitude for gluons in ${\cal N}=4$ SYM. For 
2 to 2 scattering of massless particles, there are 4 momenta, each with 
$E=|\vec{p}|$ ($k^{\mu}=(E,p^1,p^2,p^3)$). In the center
of mass frame, conservation of energy and momentum implies that they are all equal, 
$E_i=|\vec{p}_i|=k$, $i=1,2,3,4$. 
As usual, we make all momenta incoming, by changing the sign of the outgoing momenta, 
so that $\sum_i k_i=0$, and the outgoing momenta have now negative energy.
Since the two incoming spatial momenta are parallel, and the two outgoing ones are also parallel
(we are in the center of mass frame), we can arange them in a parallelogram, and define 
$k_1,k_2,k_3,k_4$ cyclically around the parallelogram. Then the Mandelstam variables are
\be
s=-(k_1+k_2)^2=-4k^2 \sin ^2 \phi/2;\;\;\;\; t=-(k_1+k_4)^2=-4k^2\cos^2\phi/2;\;\;\;\;
u=-s-t
\ee
where $\phi$ is the angle between $\vec{p}_1$ and $\vec{p}_2$, thus s and t are the 
diagonals of the parallelogram, and $s=t$ corresponds to a square.

In \cite{bds}, a conjecture was put forth for the gluon scattering amplitudes
in ${\cal N}=4$ SYM. We will 
describe it in more detail in the following section, but for 4 point amplitudes, it is
given as follows. The first observation is that the amplitude factorizes as 
\be
 {\cal A}= {\cal A}_{tree} {\cal M}(s,t)
 \ee
where ${\cal A}_{tree}$ contains all the color and polarization
factors, and ${\cal M}(s,t)$ is a common function. 
Then ${\cal M}(s,t)$ is written as 
\bea
&&{\cal M}= ({\cal A}_{div,s})^2({\cal A}_{div,t})^2 exp \{ \frac{f(\lambda)}{8} 
\ln^2\frac{s}{t}+const. \} \nonumber\\&&=
exp \{ -\frac{f(\lambda)}{8}(\ln ^2 \frac{\mu^2}{-s}+\ln^2\frac{\mu^2}{-t})
-\frac{g(\lambda)}{2} (\ln \frac{\mu^2}{-s}+\ln \frac{\mu^2}{-t})+\frac{f(\lambda)}{8}
\ln^2\frac{s}{t}+const.\}
\eea
where $f(\lambda)$ is the same function appearing in the dimension of twist two operators. 

Since the color and polarization factors factorize, we can choose any ordering of $k_1,$
$k_2,$ $k_3,$ $
k_4$ to calculate ${\cal M}(s,t)$ (choosing a different ordering will result in a different 
${\cal A}_{tree}$, but the same ${\cal M}(s,t)$). In particular, we will choose the one defined above,
with $k_1,k_2,k_3,k_4$ defined cyclically around the parallelogram of spatial momenta.

The universal function ${\cal M}(s,t)$ was obtained in \cite{am}
from an AdS space calculation as follows. 
One starts with $AdS_5$ space with the metric
\be
ds^2=R^2\frac{d\vec{x}_{3+1}^2+dz^2}{z^2}
\ee
A Gross-Mende-type calculation \cite{gm}
for the scattering of open strings dual to the gluons
shows that the amplitude is dominated by a 
classical string worldsheet with vertex operator insertions at the boundary. A `T-duality'
\be
\partial_{\alpha}y^{\mu}= i w^2(z)\epsilon_{\alpha\beta} \partial_{\beta } x^{\mu}
\ee
where neither the initial or the final coordinates are compact gives again AdS space
in coordinates
\be
ds^2=R^2\frac{dy_{\mu}dy^{\mu}+dr^2}{r^2};\;\;\;\; r=\frac{R^2}{z}
\ee

In these T-dual coordinates one obtains a classical string worldsheet ending on the 
boundary at $r=0$ on a polygon made of lighlike segments dual to the momenta, 
\be
\Delta y^{\mu}=2\pi k^{\mu}
\label{deltay}
\ee
Since $y_0$ is dual to energy, increasing $y_0$ correponds to incoming momenta and 
decreasing $y_0$ to outgoing momenta.

Then the calculation of ${\cal M}(s,t)$ in these T-dual variables is formally the same as for 
the lighlike Wilson loop, i.e.
\be
{\cal M}(s,t)=e^{iS_{string}}\sim e^{-\frac{R^2}{2\pi} A}=e^{-\frac{\sqrt{\lambda}}{2\pi}A}
\ee
where A is the area of the minimal string worldsheet, which has euclidean signature.

In a static gauge $y_1=u_1,y_2=u_2$ (where $u_1,u_2$ are worldsheet coordinates), 
the string action is 
\be
S=\frac{R^2}{2\pi}\int dy_1dy_2\frac{\sqrt{1+(\partial_i r)^2-(\partial_i y_0)^2 -
(\partial_1r\partial_2 y_0-\partial_2 r\partial_1 y_0)^2}}{r^2}
\label{action}
\ee
whereas in a conformal gauge, the action is 
\be
iS=-\frac{R^2}{2\pi} \int du_1du_2 \frac{1}{2}\frac{\partial r\partial r +\partial 
y_{\mu}\partial y^{\mu}}{r^2}
\label{confg}
\ee

The lightlike contour that the Wilson loop ends on depends on the ordering of external momenta.
As we mentioned, we can choose any ordering to calculate 
${\cal M}(s,t)$, but if we choose the ordering
where $k_1,k_2$ are incoming and $k_3,k_4$ are outgoing, the projection of the Wilson loop on 
the $y_1,y_2$ plane is singular. It is composed of 2 lines, one for the incoming momenta and
one for the outgoing ones. That means that choosing $y_1=u_1, y_2=u_2$ will be problematic. 
That is the reason that we choose to define the ordering of $k_1,k_2,k_3,k_4$ cyclically 
around the parallelogram of momenta (thus $k_1$ and $k_3$ are incoming, and $k_2$ and $k_4$ are
outgoing). 

The worldsheet corresponding to $s=t$ ends on a lightlike polygon, whose projection 
in the $y_1,y_2$ plane is a square, thus the boundary conditions are
\be
r(\pm 1, y_2)=r(y_1, \pm 1)=0,\;\;\; y_0(\pm 1, y_2)=\pm y_2;\;\;\; y_0(y_1, \pm 1)=\pm y_1
\label{bdcd}
\ee
and the solution in static gauge is 
\be
y_0(y_1, y_2)=y_1y_2,\;\;\; r(y_1, y_2)= \sqrt{(1-y_1^2)(1-y_2^2)}
\label{sgsol}
\ee
or in conformal gauge 
\be
y_1=\tanh u_1;\;\;\; y_2=\tanh u_2;\;\;\; y_0=\tanh u_1\tanh u_2;\;\;\;r=\frac{1}{
\cosh u_1\cosh u_2}
\ee
This solution turns out to be the same solution found in \cite{krucz} for a worldsheet ending on a single lightlike 
cusp (used for a lightlike Wilson loop calculation). 

The solution at $s\neq t$ is obtained by a boost with $b=v\gamma$
in the embedding coordinates of AdS, giving
\bea
&& y_1=\frac{\tanh u_1}{1+b\tanh u_1\tanh u_2};\;\;\;
y_2=\frac{\tanh u_2}{1+b\tanh u_1\tanh u_2}\nonumber\\
&& y_0=\frac{\sqrt{1+b^2}\tanh u_1\tanh u_2}{1+b\tanh u_1\tanh u_2};\;\;\;
r=\frac{1}{\cosh u_1\cosh u_2}\frac{1}{1+b\tanh u_1\tanh u_2}
\label{nonzerob}
\eea
from which one extracts (after a rescaling of momenta by $a$)
\be
s=\frac{-8a^2/(2\pi )^2}{(1-b)^2};\;\;\; t=\frac{-8a^2/(2\pi )^2}{(1+b)^2}
\label{st}
\ee
The two parameters $a$ and $b$ are enough to characterize the amplitude, which is 
a function of only $s$ and $t$. 

The action on this solution is divergent, indicative of the IR divergence of the gluon 
amplitude. To deal with it, one introduces a dimensional regularization, $D=4-2\epsilon$, 
giving the T dual metric
\be
ds^2= \sqrt{c_D\lambda_D}\left(\frac{dy^2_D+dr^2}{r^{2+\epsilon}}\right)
\label{metric}
\ee
the regularized approximate solution
\be
r_{\epsilon}\sim \sqrt{1+\epsilon/2}r_{\epsilon=0};\;\;\;\; y^{\mu}_{\epsilon}\simeq y^{\mu}_{
\epsilon=0} 
\label{approx}
\ee
and the action (using that $(\partial r\partial r+\partial y_{\mu}\partial y^{\mu})/(2r^2)|
_{\epsilon=0}=1$)
\be
S=\frac{\sqrt{\lambda_Dc_D}}{2\pi}\int\frac{{\cal L}_{\epsilon=0}}{r^{\epsilon}}
=i\frac{\sqrt{\lambda_Dc_D}}{2\pi}\int_{-\infty}^{+\infty} du_1du_2
r_{\epsilon=0}^{-\epsilon}
[1+\frac{\epsilon}{2}(\frac{\partial r\partial r}{2r^2}|_{\epsilon=0}-1)
-\frac{\epsilon^2}{4}(\frac{\partial r\partial r}{2r^2}|_{\epsilon=0}-1)-\frac{\epsilon^2}{4}]
\label{regaction}
\ee

The AdS calculation then reproduces the BDS result, giving the values of $f(\lambda)$ 
and $g(\lambda)$ at strong coupling
\be
f=\frac{\sqrt{\lambda}}{\pi};\;\;\;\; g=\frac{\sqrt{\lambda}}{2\pi}(1-\ln 2)
\ee

\section{Six-point scattering amplitudes}

In this section we present six-point scattering amplitudes at strong coupling --- these 
solutions did not appear previously in the literature. We start in the first subsection with a 
review of 
BDS conjecture --- following \cite{dks, bht, mmt} we also present a pictorial representation at weak 
coupling for the finite part of a six-point amplitude. Then, in the next subsection we explictly construct 
and discuss in detail our new lightlike Wilson loop solutions in AdS.

\subsection{Six-point functions: field theory}

Bern, Dixon and Smirnov \cite{bds} have conjectured more general formulas for the gluon 
amplitudes, applicable to any n-point function. 

The first observation is that the same factorization of color and polarization factors 
applies for any $n$-point amplitude, and we have
\be
{\cal A}_n={\cal A}_n^{tree} {\cal M}_n(\epsilon)
\ee
where ${\cal M}_n$ only depends on momentum invariants and the dependence on $\epsilon$ indicates that we use the dimensional regularization. The supersymmetry constraints the kinematic dependent part to take a nice {\it exponential} form --- specifically, ${\cal M}_n(\epsilon)$ can be factorized in 
an infrared divergent part, a finite part, and a coupling-dependent constant: 
\bea
{\cal M}_n(\epsilon)&=&{\cal M}^{IR}_n(\epsilon)F_n(\epsilon)C(\lambda)=\exp\left[
\sum_{l=1}^{\infty}a^l f^{(l)}(\epsilon)\hat{I}^{(1)}_n(l\epsilon)
\right]
\tilde{h}_n(\epsilon)\nonumber\\
&=&\exp\left[
\sum_{l=1}^{\infty}a^l f^{l}(\epsilon)\hat{I}^{(1)}_n(l\epsilon)+\sum_{l=1}^{\infty}
a^lf^{(l)}(\epsilon)F_n^{(1)}(l\epsilon)+\sum_{l=1}^{\infty}a^lh_n^{(l)}(\epsilon)
\right]
\eea
The constant $a$ is a function of 't Hooft coupling, $\lambda$, and the dimensional regularization 
parameter $\epsilon$:
\be
a=\lambda (4\pi e^{-\gamma})^{-\epsilon}
\ee
where $\gamma$ is the Euler's constant. In the limit $\epsilon\rightarrow 0$, the 
constant $a$ becomes 't Hooft coupling $\lambda$. The functions $f^{(l)}(\epsilon)$ have 
a perturbative expansion
\be
f^{(l)}(\epsilon)=f_0^{(l)}+f_1^{(l)}\epsilon+f_2^{(l)}\epsilon^2
\ee
where the first term in expansion, $f_0^{(l)}$, is related to the cusp anomalous dimension 
for an $l$-loop. Here $M_n^{(1)}(\epsilon)=I_n^{(1)}(\epsilon)+F_n^{(1)}(\epsilon)$ is the 
1-loop amplitude divided by the tree amplitude, thus up to constants and functions of $\lambda
$ the amplitude is the exponential of the 1-loop amplitude. 
The IR divergent part, ${\cal M}^{IR}_n(\epsilon)$, is controlled by 
the factor $\hat{I}_n^{(1)}(\epsilon)$ that contains $1/\epsilon^2$ IR divergencies. 
The finite part $F_n(\epsilon)$ that is controlled by the factor $F_n^{(1)}(\epsilon)$ is known as 
the finite remainder (it is finite as $\epsilon\rightarrow 0$), and $h_n^{(l)}(\epsilon)$ are 
constant factors which do not depend on kinematics. 

The divergent factor is
\be
\hat{I}^{(1)}_n(\epsilon)=-\frac{1}{2}\frac{1}{ \epsilon^2}
\sum_{i=1}^{n} \left( \frac{\mu^2}{ -s_{i,i+1}}\right)^{\epsilon}=-\frac{1}{2\epsilon^2}
\sum_{i=1}^n[1+\epsilon \ln \left( \frac{\mu^2}{ -s_{i,i+1}}\right)
+\frac{\epsilon^2}{2} \Bigl(\ln\left( \frac{\mu^2}{
			       -s_{i,i+1}}\right)\Bigr)^2
+\cdots]
\ee
where $s_{i,i+1}\equiv (k_i+k_{i+1})^2$ are Madelstam variables for any neighboring pair of gluons and $\mu$ 
is the renormalization scale parameter.
Then the amplitude is expanded in $\epsilon$ as
\begin{eqnarray}
\ln {\cal M}_n(\epsilon)&=&\frac{A_2}{ \epsilon^2}+\frac{A_1}{ \epsilon}+A_0
\CR
&&
-\frac{1}{4} \sum_{i=1}^{n}
\sum_{l=1}^{\infty}f^{(l)}_0a^l 
\Bigl(\ln\left( \frac{\mu^2}{
-s_{i,i+1}}\right)\Bigr)^2
-\frac{1}{2}  \sum_{i=1}^{n}
\sum_{l=1}^{\infty}\frac{f^{(l)}_1}{ l}a^l
\ln\left( \frac{\mu^2}{ -s_{i,i+1}}\right)
\CR
&& +\sum_{l=1}^{\infty}f^{(l)}_0 a^l F_n^{(1)}(0)+O(\epsilon)
\end{eqnarray}
where
\begin{eqnarray}
 A_2&=&-\frac{n}{2}\sum_{l=1}^{\infty}\frac{f_0^{(l)}}{ l^2}a^l 
\CR
A_1&=&-\frac{n}{2} \sum_{l=1}^{\infty}\frac{1}{ l^2} f_1^{(l)} a^l
-\frac{1}{2}\sum_{l=1}^{\infty} \frac{f^{(l)}_0}{ l} a^l
\sum_{i=1}^n \ln\left( \frac{\mu^2}{
-s_{i,i+1}}\right)\CR
A_0&=& -\frac{n}{2}\sum_{l=1}^{\infty}\frac{f_2^{(l)}}{ l^2} a^l
\end{eqnarray}

Following \cite{buch}, we define
\begin{equation}
 f(\lambda)=4 \sum_{l=1}^{\infty}f^{(l)}_0 a^l;\;\;\;
 g(\lambda)=2 \sum_{l=1}^{\infty}\frac{f^{(l)}_1}{ l}a^l
\end{equation}
where $f(\lambda)$ and $g(\lambda)$ are the same functions as defined for the 4-point 
function. In the limit $\epsilon\rightarrow 0$, $f(\lambda)=4 \sum_{l=1}^{\infty}f^{(l)}_0\lambda^l$ is the all-loop cusp anomalous dimension that appears in the dimension of twist two operators. We then 
obtain for the finite (in $\epsilon$, but still IR divergent in $\mu$) part of the 
amplitude
\begin{eqnarray}
\left. \ln {\cal M}_n\right|_{\epsilon^0}
&=& A_0-\frac{1}{16}f(\lambda)
\sum_{i=1}^{n}\Bigl(\ln\left( \frac{\mu^2}{
-s_{i,i+1}}\right)\Bigr)^2
-\frac{g(\lambda)}{4}
\sum_{i=1}^{n} 
\ln\left( \frac{\mu^2}{ -s_{i,i+1}}\right)
+\frac{f(\lambda)}{4}F_{n}^{(1)}(0)
\CR
\label{eq:npt}
\end{eqnarray}

Finally, the finite remainder for $n>4$ is given by (for $n=4$, $F_n^{(1)}(0)=1/2 \ln^2 s/t$):
\be
\frac{f(\lambda)}{4}F_n^{(1)}(0)=\frac{f(\lambda)}{4}
\frac{1}{2}\sum_{i=1}^ng_{n,i}
\ee
where the functions $g_{n,i}$ contain dilogarithms and squares of ordinary logarithms
\begin{equation}
 g_{n,i}=-\sum_{r=2}^{[n/2]-1}
\ln\left(\frac{-t^{[r]}_{i}}{ -t^{[r+1]}_{i}}\right)
\ln\left(\frac{-t^{[r]}_{i+1}}{ -t^{[r+1]}_{i}}\right)
+D_{n,i}+L_{n,i}+\frac{3}{2}\zeta_{2}
\label{eq:bds1}
\end{equation}
Here we used the `generalized' Mandelstam variables $t_i^{[r]}\equiv (k_i+...+k_{i+r-1})^2$ (mod n for 
the index $i$). The others terms are given by

$\bullet n=2m+1$

\begin{eqnarray}
 D_{2m+1}&=&
-\sum_{r=2}^{m-1}{\rm Li}_2\left(1-\frac{t^{[r]}_{i} t^{[r+2]}_{i-1}}{ 
t^{[r+1]}_i t^{[r+1]}_{i-1}}\right)\\
L_{2m+1}&=& -\frac{1}{2}
\ln\left(\frac{-t^{[m]}_{i}}{ -t^{[m]}_{i+m+1}}\right)
\ln\left(\frac{-t^{[m]}_{i+1}}{ -t^{[m]}_{i+m}}\right)
\end{eqnarray}

$\bullet n=2m$

\begin{eqnarray}
 D_{2m}&=&
-\sum_{r=2}^{m-2}{\rm Li}_2\left(1-\frac{t^{[r]}_{i} t^{[r+2]}_{i-1}}{ 
t^{[r+1]}_i t^{[r+1]}_{i-1}}\right)
-\frac{1}{2}{\rm Li}_2\left(1-\frac{t^{[m-1]}_{i} t^{[m+1]}_{i-1}}{ 
t^{[m]}_i t^{[m]}_{i-1}}\right)
\\
L_{2m}&=& -\frac{1}{4}
\ln\left(\frac{-t^{[m]}_{i}}{ -t^{[m]}_{i+m+1}}\right)
\ln\left(\frac{-t^{[m]}_{i+1}}{ -t^{[m]}_{i+m}}\right)
\end{eqnarray}
and some useful dilogarithmic relations are
\begin{equation}
Li_2(z)=\sum_{k=1}^{\infty}\frac{z^k}{k^2}, \,\,\,\,\, Li_2(0)=0, \,\,\,\,\, Li_2(1)=\zeta_{2}=\frac{\pi^2}{6}
\end{equation}

We can now use the input of the 4-point amplitude AdS calculation of Alday and Maldacena,
and substitute the large $\lambda$ value of $f(\lambda)$ and $g(\lambda)$ in the above
formulas. Since
\bea
&&f(\lambda) =\frac{\sqrt{\lambda}}{\pi}=4\sum_{l\ge 1}a^l f_0^{(l)}\nonumber\\
&&g(\lambda)=\frac{\sqrt{\lambda}}{2\pi}(1-\ln 2)=2\sum_{l\geq 1}\frac{a^lf_1^{(l)}}{l}
\eea
by acting with $(\lambda d/d\lambda)^{-1}$ once on g and once and twice on f, we get
\bea
&&\frac{1}{4}f^{-1}(\lambda)\equiv \sum_{l\geq 1}\frac{a^l f_0^{(l)}}{l}=\frac{\sqrt{\lambda}}{2\pi};\;\;\;
\frac{1}{4}f^{-2}(\lambda)\equiv \sum_{l\geq 1}\frac{a^lf_0^{(l)}}{l^2}=\frac{\sqrt{\lambda}}{\pi}\nonumber\\
&&\frac{1}{2}g^{-1}(\lambda)\equiv\sum_{l\geq 1}\frac{a^l f_1^{(l)}}{l^2}=\frac{\sqrt{\lambda}}{2\pi}(1-\ln 2)
\eea
Substituting these functions in the amplitude, we get at large 
coupling (ignoring terms $O(\epsilon)$)
\bea
&&\ln {\cal M}_n =A_0 - \frac{n\sqrt{\lambda}}{2\pi}\frac{1}{\epsilon^2}-\frac{1}{\epsilon}
\left[\frac{n\sqrt{\lambda}}{4\pi}(1-\ln 2)+\frac{\sqrt{\lambda}}{4\pi}\sum_{i=1}^n\ln 
\frac{\mu^2}{-s_{i,i+1}}\right]\nonumber\\
&&-\frac{\sqrt{\lambda}}{16\pi}
\sum_{i=1}^{n}\ln^2\left( \frac{\mu^2}{
-s_{i,i+1}}\right)
-\frac{\sqrt{\lambda}}{8\pi}(1-\ln 2)
\sum_{i=1}^{n} 
\ln\left( \frac{\mu^2}{ -s_{i,i+1}}\right)
+\frac{\sqrt{\lambda}}{4\pi}F_{n}^{(1)}(0)
\label{divamp}
\eea

For $n=6$ we have $s_{i,i+1}\equiv t_i^{[2]}$,
$t_i^{[3]}=t_{i+3}^{[3]}$ and $t_i^{[4]}=t_{i-2}^{[2]}$ (due to momentum 
conservation) and then the finite part is given by
\be
F_6^{(1)}(0)=-\frac{1}{2}\sum_{i=1}^6\left[
\ln\frac{t_i^{[2]}}{t_{i}^{[3]}}\ln\frac{t_{i+1}^{[3]}}
{t_i^{[3]}}+\frac{1}{2}{\rm Li}_2\left(1-\frac{t_i^{[2]}t_{i-3}^{[2]}}{t_i^{[3]}t_{i-1}^{[3]}}
\right)-\frac{1}{4}\ln^2\frac{t_i^{[3]}}{t_{i+1}^{[3]}}\right]
\ee

Since $M_6^{(1)}(\epsilon)=I_6^{(1)}(\epsilon)+F_6^{(1)}(\epsilon)$ is the 1-loop amplitude, 
this formula has an interesting representation. Indeed, the 1-loop amplitude can be
written as a sum over box integrals. 
A nice pictorial representation of this decomposition is to form "clusters"  from 
external momenta of the 1-loop diagrams and diagonals of the same \cite{mmt}. 
The diagonals are then replaced by a partial sum of external momenta and so can be 
interpreted as off-shell momenta. The clusters with two opposite momenta off-shell and the other two on-shell are called two-mass easy box functions and are usually denoted by $F^{\rm 2m\,e}$ \cite{Bern:1993kr}. The clusters with three or four null 
(on-shell) momenta correspond to one-mass and zero-mass boxes.

For a 6-point amplitude there are two kinds of 4-clusters: the degenerate one 
($F^{2me}_{1;i}$) formed from three on-shell external momenta and one off-shell momentum (one diagonal) and the other one ($F^{2me}_{2;i}$)
formed from two on-shell external momenta and two off-shell momenta (two diagonals). 
Thus, we obtain \cite{bddk,bst} (see also \cite{ftt}):
\begin{equation}
M_{6}^{(1)}   \ = \  \frac{\Gamma (1 + \epsilon) \Gamma^2 ( 1 -  \epsilon)} {  (4\pi)^{2- \epsilon}
\Gamma(1 - 2 \epsilon)}\sum_{i=1}^{6} \sum_{r=1}^{2}
\Bigl(1 - \frac{1}{2} \delta_{2, r}\Bigr)\,
F_{r;i}^{\rm 2m\,e} (p, q, P, Q)
\end{equation}
where $p=p_{i-1}$, $q=p_{i+r}$, $P=p_i + \cdots +p_{i+r-1}$, and $p+q+P+Q=0$.

An useful form (all-orders in $\epsilon$) of the two-mass easy box function is given by \cite{ftt} 
\bea
&&
F^{\rm 2me} (s, t, P^2, Q^2) =
-\frac{1}{ \e^2}
\left[
 \Big( \frac{-s}{ \mu^2} \Big)^{-\e} \,
\, + \,
\Big( \frac{-t }{ \mu^2} \Big)^{-\e}
\, - \, 
\Big( \frac{-P^2 }{ \mu^2} \Big)^{-\e}
\, - \, 
 \Big( \frac{-Q^2 }{ \mu^2} \Big)^{-\e} \,
\right. 
\\ 
&& \hspace{-0.3cm} \left. + \, 
\Big( \frac{a \m^{2} }{ 1-aP^2} \Big)^{\e} \,
\mbox{}_{2}F_1 \left( \e, \e, 1+ \e, \frac{1 }{ 1 - a P^2 }\right)
\, + \,
\Big( \frac{a \m^{2} }{ 1-aQ^2} \Big)^{\e} \,
\mbox{}_{2}F_1 \left( \e, \e, 1+ \e, \frac{1 }{ 1 - a Q^2 }\right)
\right.
\nonumber 
\\ [6pt] \cr
&& \hspace{-0.3cm}  - \,
\left.
\Big( \frac{a \m^{2} }{ 1-as} \Big)^{\e} \,
\mbox{}_{2}F_1 \left( \e, \e, 1+ \e, \frac{1 }{ 1 - a s }\right)
\, - \,
 \Big( \frac{a \m^{2} }{ 1-at} \Big)^{\e} \,
\mbox{}_{2}F_1 \left( \e, \e, 1+ \e, \frac{1 }{ 1 - a t }\right)
\right]
\ . 
\nonumber
\eea

where

\begin{equation} 
a \ = \
\frac{P^2+Q^2-s-t}{P^2Q^2-st} 
\end{equation}

and $s := (P+p)^2$,  $t := (P+q)^2 $.

The first line is the divergent part of the two-mass easy box function that matches the divergent part of 
the on-shell up to a factor of 2 \cite{dks}. After taking the limit $\epsilon\rightarrow 0$, the finite part contains only the following 
dilogarithms  \cite{Duplancic:2000sk, bst}  
\begin{equation} 
{\rm Li}(1-aP^2)\, + \, {\rm Li}(1-aQ^2)  \, -\,  {\rm Li}(1-as)\,  -\,  
{\rm Li}(1-at)
\end{equation}

The degenerate cluster (one-mass function)
does not contribute to the dilogarithmic part of the BDS formula and since the
4- and 5-point amplitudes only contain this cluster, these amplitudes 
do not contain dilogarithmic terms. 

The duality between lightlike Wilson loops and gluon amplitudes holds 
also in the weak coupling limit. Thus, 
to make connection with the Wilson loop computations at strong coupling it would 
be interesting to understand 
the MHV amplitudes from a Wilson loop computation at weak coupling. There are two one-loop 
corrections to the Wilson loop. When the gluon stretches between two lightlike momenta meeting 
at a cusp there is a contribution to 
the infrared divergent part of the amplitude. When the gluon stretches between two non-adjacent 
segments there is a contribution to the finite part. 

We will see in the next sections that 
the AdS-CFT dual amplitudes have extra restrictions, that should correspond to restrictions
on the allowed Feynman diagrams in the amplitude. Clearly, these 
conditions can modify the above cluster decomposition for ${\cal M}_n$.

\subsection{Six-point amplitudes: AdS}

The Alday-Maldacena solution for the Wilson loop ending on a square in $y_1,y_2$ is given 
in (\ref{sgsol}), and is a solution of the action (\ref{action}) with boundary conditions
(\ref{bdcd}). We use the symmetries of the action to construct new simple solutions. Thus, 
by cutting and gluing these solutions and a careful consideration of the boundary 
conditions we construct 6-point function solutions of the same action. 

First, by noticing that we can change the sign of $y_0$ in (\ref{action}), we can 
construct the solution
\be
y_0(y_1,y_2)=y_1 |y_2|,\;\;\;\;\;
r(y_1,y_2)=\sqrt{(1-y_1^2)(1-y_2^2)}\label{sol2_r}
\ee
(solution 2 in the following) and also a `composed' solution  (solution 1 in the following)
\bea
&&y_0(y_1,y_2)=\frac{1}{2}(|y_1 y_2|+y_1 y_2-|y_1|y_2+y_1|y_2|)\label{sol1_y0},\nonumber\\
&&r(y_1,y_2)=\sqrt{(1-y_1^2)(1-y_2^2)}\label{sol1_r}
\eea

\begin{figure}[bthp]
\begin{center}\includegraphics{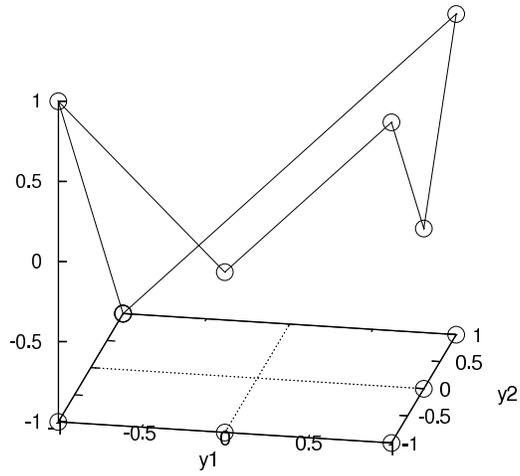}\end{center}
\caption{Solution 1: $y_0=1/2(|y_1y_2|+y_1y_2-|y_1|y_2+y_1|y_2|)$}\label{sol1}
\end{figure}

\begin{figure}[bthp]
\begin{center}\includegraphics{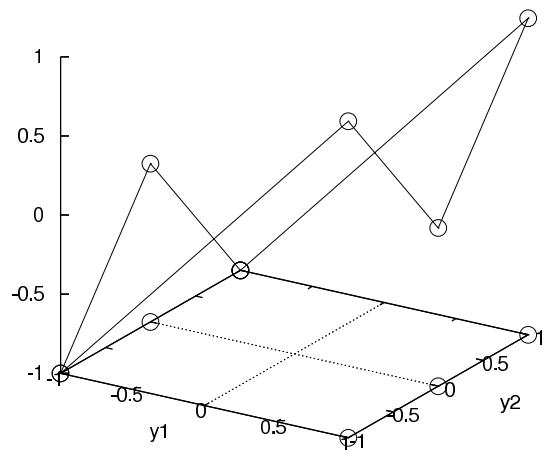}\end{center}
\caption{Solution 2: $y_0=y_1|y_2|$}\label{sol2}
\end{figure}

The boundary conditions for these solutions are drawn in Fig.\ref{sol1} and Fig.\ref{sol2},
from where we see that they indeed are 6-point functions.

Another 6-point function solution is found by replacing $y_2\rightarrow -2+|y_2|$ in the Alday-Maldacena solution (and shifting $y_0$ for convenience), i.e.
\be
y_0-2=(-2+|y_2|)y_1;\;\;\;\;\; r^2=(1-y_1^2)(1-(-2+|y_2|)^2)
\ee
which again takes advantage of the symmetries of the action and gluing. One can check that the
external (incoming and outgoing) momenta are the same for this solution as for the $y_0=y_1
|y_2|$ solution, just with a different colour ordering. We argued that we can choose any colour
ordering to calculate ${\cal M}(s,t)$ and we will get the same function. Indeed, since 
these 2 solutions have the same action (they were obtained by symmetries and cutting and 
gluing), they do give the same result. The external momenta will be in principle different at
nonzero $b$, but we will not analyze this solution further. 

Note that the new solutions are not guaranteed to be valid on the lines where we glue 
them. We will come back to this point at the end of this section, but for the moment we 
will ignore it.

At this point the new solutions are just a trivial redefining of the old one, but we now
need to find the solution for varying external momenta. In the case of the 4-point function,
there were only 2 invariant variables, $s$ and $t$, and consequently we could obtain them from 
a boost parameter $b$ in the auxiliary embedding coordinate of AdS and an overall scaling
by $a$. For the 6-point function, these two parameters are not enough, since we have more
external momenta. In fact there are 8 variables: 6 momenta, minus the center of mass
momentum, minus the one momentum given by momentum conservation give 4 momenta. The mass 
shell conditions of the 4 momenta, spatial rotations, and the mass shell condition of the 
sum of 5 momenta reduce it to 8 variables.

But what we can do is to make the same transformation as for the 4-point function, depending
on parameters $a$ and $b=v\gamma$. We go to the AdS embedding coordinates
\bea
&&Y^{\mu}=\frac{y^\mu}{r} \quad (\mu=0,\cdots, 3),\nonumber\\
&&Y_{-1}+Y_{4}=\frac{1}{r}, \qquad Y_{-1}-Y_4=\frac{r^2+y_\mu y^\mu}{r}.
\eea
and perform a Lorentz boost in the 04 plane, 
\be
 \begin{pmatrix}
  Y'^0\\Y'^4
 \end{pmatrix}
=
\begin{pmatrix}
\gamma &  v \gamma \\ v\gamma & \gamma
\end{pmatrix}
 \begin{pmatrix}
  Y^0\\Y^4
 \end{pmatrix},
\ee
with $\gamma=1/\sqrt{1-v^2}$ and a rescaling by $a$, after which the solution becomes
(using that $Y_4\sim 1-r^2-y_{\mu}y^{\mu}=0$)
\be
 r'=\frac{a r(y_1,y_2)}{1+b y_0(y_1,y_2)},\;\;\;\;\;
y'_0=\frac{a\sqrt{1+b^2}y_0(y_1,y_2)}{1+b y_0(y_1,y_2)},\;\;\;\;\;
y'_i=\frac{a\sqrt{1+b^2}y_i(y_1,y_2)}{1+b y_0(y_1,y_2)}
\ee

\begin{figure}
\begin{minipage}{7cm}\vspace{-.7cm}\includegraphics{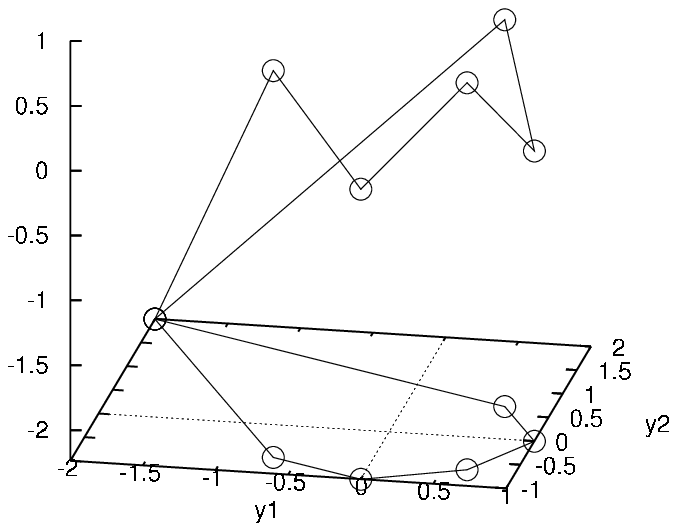}\end{minipage}
\begin{minipage}{5cm}\includegraphics{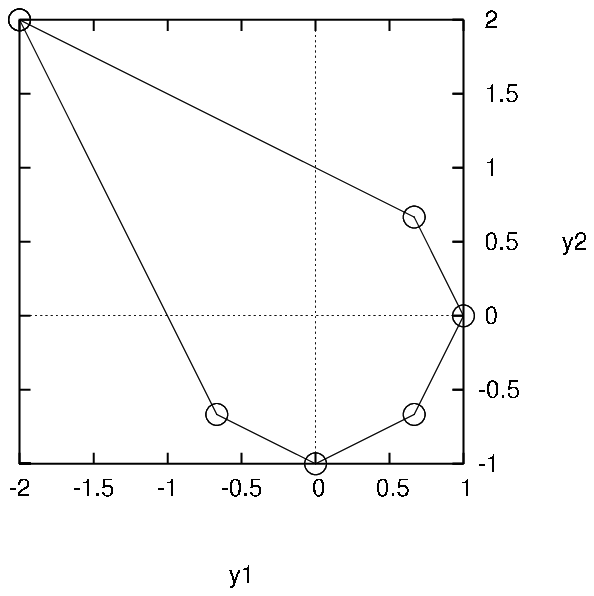}\end{minipage}
\caption{Configuration after the Lorentz boost
in the 04 plane for solution 1, $y_0$ $=1/2(|y_1y_2|+y_1y_2-|y_1|y_2+y_1y_2)$ with a=1, b=0.5.}
\label{afterBoost1}
\end{figure}

\begin{figure}
\begin{minipage}{7cm}\vspace{-.7cm}\includegraphics{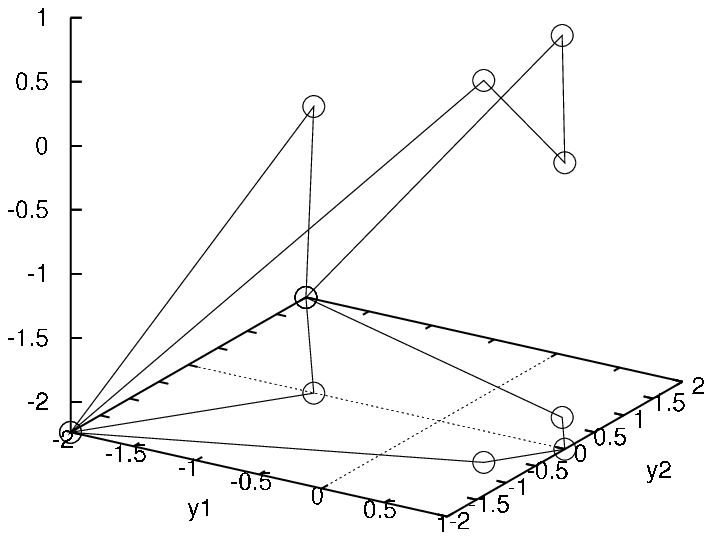}\end{minipage}
\begin{minipage}{5cm}\includegraphics{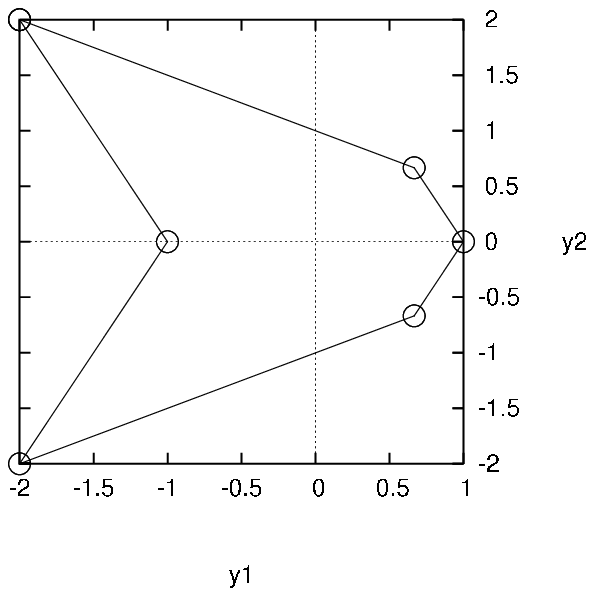}\end{minipage}
\caption{Configuration after the Lorentz boost
in the 04 plane for solution 2, $y_0=y_1|y_2|$ with a=1, b=0.5.}
\label{afterBoost2}
\end{figure}

The boundaries of the boosted solutions (\ref{sol1_r}) and (\ref{sol2_r}) are depicted 
in Fig.\ref{afterBoost1} and Fig.\ref{afterBoost2}.
In conformal gauge, these solutions are 
\bea
&&r=\frac{a}{\cosh u_1 \cosh u_2\pm b \sinh u_1 \sinh u_2},\quad
y_0=\frac{\pm a\sqrt{1+b^2}\sinh u_1\sinh u_2}
{\cosh u_1 \cosh u_2\pm b \sinh u_1 \sinh u_2}\nonumber\\&&
y_1=\frac{a\sinh u_1\cosh u_2}
{\cosh u_1 \cosh u_2\pm b \sinh u_1 \sinh u_2},\quad
y_2=\frac{a\cosh u_1\sinh u_2}
{\cosh u_1 \cosh u_2\pm b \sinh u_1 \sinh u_2}
\label{boosted}
\eea
where $-/+$ corresponds to $\{u_1>0,u_2<0\}$/(others)
for the solution $y_0(y_1,y_2)=1/2(|y_1y_2|+y_1y_2-|y_1|y_2+y_1|y_2|)$,
and to $\{u_2>0\}$/$\{u_2<0\}$ for 
the solution $y_0(y_1,y_2)=y_1|y_2|$.

We can read off the external momenta corresponding to these solutions by going to the 
boundary and defining $k^i=(\Delta  y^{(i)}_1, \Delta y^{(i)}_2,\Delta y^{(i)}_0)$,
where $\Delta y^{(i)}_{\mu}\equiv y'_{\mu}(P_{i+1})-y'_\mu(P_{i})$. Here $P_i$ are the 
vertices of the boundary Wilson line, specifically
$P_1$, $P_2$, $P_3$, $P_4$, $P_5$ and $P_6$
correspond to $(y_1,y_2)=(-1,1)$, $(-1,-1)$, $(0,-1)$, $(1,-1)$, $(1,0)$ and
$(1,1)$ 
in the original coordinate before the boost,
for the solution $y_0(y_1,y_2)=1/2(|y_1y_2|+y_1y_2-|y_1|y_2+y_1|y_2|)$,
and to $(y_1,y_2)=(-1,1)$, $(-1,0)$, $(-1,-1)$, $(1,-1)$, $(1,0)$ and
$(1,1)$ for the solution $y_0(y_1,y_2)=y_1|y_2|$. We then obtain the momenta
\bea
k_1=\left(\frac{2 a b}{1-b^2},-\frac{2 a}{1-b^2},\frac{2 a\sqrt{1+b^2}}{1-b^2}\right),&&
k_2=\left(\frac{a}{1+b},-\frac{a b}{1+b},-\frac{a
 \sqrt{1+b^2}}{1+b}\right),
\nonumber\\
k_3=\left(\frac{a}{1+b},\frac{a b}{1+b},\frac{a
 \sqrt{1+b^2}}{1+b}\right),&&
k_4=\left(\frac{a b}{1+b},\frac{a}{1+b},-\frac{a
 \sqrt{1+b^2}}{1+b}\right),
\nonumber\\
k_5=\left(-\frac{a b}{1+b},\frac{a}{1+b},\frac{a
 \sqrt{1+b^2}}{1+b}\right),&&
k_6=\left(-\frac{2 a}{1-b^2},\frac{2 a b}{1-b^2},-\frac{2 a\sqrt{1+b^2}}{1-b^2}\right)
\label{kone}
\eea
for the solution $y_0(y_1,y_2)=1/2(|y_1y_2|+y_1y_2-|y_1|y_2+y_1|y_2|)$,
and 
\bea
k_1=\left(\frac{a b}{1-b},\frac{a}{b-1},\frac{a \sqrt{b^2+1}}{1-b}\right),&&
k_2=\left(\frac{a b}{b-1},\frac{a}{b-1},\frac{a
 \sqrt{b^2+1}}{b-1}\right)
\nonumber\\
k_3=\left(-\frac{2 a}{b^2-1},-\frac{2 a b}{b^2-1},-\frac{2 a\sqrt{b^2+1}}{b^2-1}\right),&&
k_4=\left(\frac{a b}{b+1},\frac{a}{b+1},-\frac{a
 \sqrt{b^2+1}}{b+1}\right)\nonumber\\
k_5=\left(-\frac{a b}{b+1},\frac{a}{b+1},\frac{a \sqrt{b^2+1}}{b+1}\right)&&
k_6=\left(\frac{2 a}{b^2-1},-\frac{2 a b}{b^2-1},\frac{2 a\sqrt{b^2+1}}{b^2-1}\right)
\label{ktwo}
\eea
for the solution $y_0(y_1,y_2)=y_1|y_2|$.

We note that the sum of the incoming momenta (if $b<1$), $k_1+k_3+k_5$, is 
$a/(1-b^2)(1+b^2, -(1+b^2),2(2-b)\sqrt{1+b^2})$ for solution 1 and
$2a/(1-b^2)(1+b^2,0,2\sqrt{1+b^2})$ for solution 2, so both are not in the center of mass
frame.

We now calculate the AdS amplitude as the exponential of the string action. Since we still 
have $(\partial r\partial r+\partial y_{\mu}\partial y^{\mu})/(2r^2)|
_{\epsilon=0}=1$ for the new solutions, the dimensionally regularized action 
on the solution is still (\ref{regaction}). The dimensionally regularized solution is again
(\ref{approx}), i.e.
\be
r_{\epsilon}\sim \sqrt{1+\epsilon/2}r_{\epsilon=0};\;\;\;\; y^{\mu}_{\epsilon}\simeq y^{\mu}_{
\epsilon=0} 
\ee

The leading term in (\ref{regaction}) is then 
\be
-iS=\frac{\sqrt{\lambda_Dc_D}}{2\pi}\int_{-\infty}^{\infty}
du_1 du_2
(\cosh u_1 \cosh u_2 + \beta \sinh u_1 \sinh u_2)^\epsilon
\ee
where $\beta = \mp b$ for $\{u_1>0,u_2<0\}$/(others)
for the solution $y_0(y_1,y_2)=1/2(|y_1y_2|+y_1y_2-|y_1|y_2+y_1|y_2|)$,
and $\{u_2>0\}$/$\{u_2<0\}$  for 
the solution $y_0(y_1,y_2)=y_1|y_2|$. We have calculated the subleading terms and they 
give constant finite contributions as in the \cite{am} case, therefore we will drop 
them (since we are not considering these constant terms).

The details of the evaluation of the integral are given in the Appendix. For the 
solution $y_0(y_1,y_2)=1/2(|y_1y_2|+y_1y_2-|y_1|y_2+y_1|y_2|)$ we obtain
\bea
&&I=\int_{-\infty}^{\infty}
du_1 du_2
(\cosh u_1 \cosh u_2 + \beta \sinh u_1 \sinh u_2)^\epsilon\nonumber\\
&&=\frac{\pi \Gamma[-\frac{\epsilon}{2}]^2}{\Gamma[\frac{1-\epsilon}{2}]^2}
~_2F_1(\frac{1}{2},-\frac{\epsilon}{2},\frac{1-\epsilon}{2};b^2)+
\frac{2 b}{\epsilon}
~_3F_2(1,1,\frac{1-\epsilon}{2};\frac{3}{2},1-\frac{\epsilon}{2};b^2)\label{result2}
\eea
where the first term in the last line corresponds to the 4-point function result, and 
the second is a new contribution. For the solution $y_0(y_1,y_2)=y_1|y_2|)$, we obtain 
only the first term, thus the same result as for the 4-point function. Using the 
expansion of the hypergeometric functions,
\bea
&&~_2F_1(\frac{1}{2},-\frac{\epsilon}{2},\frac{1-\epsilon}{2};b^2)
=1+\frac{1}{2}\ln(1-b^2)\epsilon
+\frac{1}{2}\ln(1-b)\ln(1+b)\epsilon^2+{\cal O}(\epsilon^3)\nonumber\\
&&~_3F_2(1,1,\frac{1-\epsilon}{2};\frac{3}{2},1-\frac{\epsilon}{2};b^2)\nonumber\\
&&=\frac{1}{2b}\ln\left(\frac{1+b}{1-b}\right)
+\frac{1}{2b}\left\{
-\ln 2\ln\left(\frac{1+b}{1-b}\right)
-\text{Li}_2\left(\frac{1-b}{2}\right)+\text{Li}_2\left(\frac{1+b}{2}\right)
\right\}\epsilon
\eea
we obtain the AdS result
\bea
&& -\frac{\sqrt{\lambda}}{2\pi}(2\pi^2\frac{\mu^2}{4a^2})^{\epsilon/2}\left[\frac{4}{\epsilon
^2}+\frac{2}{\epsilon}\ln (1-b^2)+\frac{2}{\epsilon}(1-\ln 2)\right.\nonumber\\&&\left.
+(1-\ln 2)\ln (1-b^2)
+2\ln (1-b)\ln (1+b)\right.\nonumber\\&&
\left.+\frac{1}{\epsilon}\ln\frac{1+b}{1-b}+\frac{1+\ln 2}{2}\ln \frac
{1+b}{1-b}-{\rm Li}_2\left(\frac{1-b}{2}\right)+{\rm Li}_2\left(\frac{1+b}{2}\right)\right]
\label{adsresult}
\eea
for the $y_0(y_1,y_2)=1/2(|y_1y_2|+y_1y_2-|y_1|y_2+y_1|y_2|)$ solution, where the first 
two lines are the 4-point function result and the last line is the extra term. 
For the $y_0(y_1,y_2)=y_1|y_2|)$ solution, we have only the first two lines, i.e. the 4-point function result. Note that the normalization of $\mu^2$ by $2\pi^2$ is the same as in \cite{am}
(part of it is a $(2\pi)^2$ in $t^{[2]}_i$'s in (\ref{deltay}) and (\ref{st}), and also a 
factor of 2). The contributions to the action  from the higher order $\epsilon$ terms in 
(\ref{regaction})
is evaluated in a similar way. For the solution 1,  $+1$ is added in the square bracket in 
\ref{adsresult} which is same as the 4-point case. For the solution 2, $1-b$ is added.

We will now apply our 6-point function field theory formulas for the momenta in 
(\ref{kone}) and (\ref{ktwo}) and compare with the AdS results.
For these momenta, the relevant $t_i^{[r]}$ variables are given by 
\bea
&&t^{[2]}_1=\frac{4 a^2}{1-b},\;\;
t^{[2]}_2=\frac{4 a^2}{(b+1)^2},\;\;
t^{[2]}_3=2 a^2,\;\;
t^{[2]}_4=\frac{4 a^2}{(b+1)^2},\;\;
t^{[2]}_5=\frac{4 a^2}{1-b},\;\;
t^{[2]}_6=\frac{8 a^2}{(b+1)^2}\nonumber\\
&&t^{[3]}_1=\frac{4 a^2}{1-b^2},\;\;
t^{[3]}_2=\frac{4 a^2}{b+1},\;\;
t^{[3]}_3=\frac{4 a^2}{b+1},\;\;
t^{[3]}_4=\frac{4 a^2}{1-b^2},\;\;
t^{[3]}_5=\frac{4 a^2}{b+1},\;
t^{[3]}_6=\frac{4 a^2}{b+1}
\eea
 for solution 1 ($y_0(y_1,y_2)=1/2(|y_1y_2|+y_1y_2-|y_1|y_2+y_1|y_2|)$) and
\bea
&&t^{[2]}_1=\frac{4 a^2}{(1-b)^2},\;\;
t^{[2]}_2=\frac{4 a^2}{b+1},\;\;
t^{[2]}_3=\frac{4 a^2}{1-b},\;\;
t^{[2]}_4=\frac{4 a^2}{(b+1)^2},\;\;
t^{[2]}_5=\frac{4 a^2}{1-b},\;\;
t^{[2]}_6=\frac{4 a^2}{b+1},\nonumber\\
&&t^{[3]}_1=\frac{4 a^2}{1-b^2},\;\;
t^{[3]}_2=4 a^2,\;\;
t^{[3]}_3=\frac{4 a^2}{1-b^2},\;\;
t^{[3]}_4=\frac{4 a^2}{1-b^2},\;\;
t^{[3]}_5=4 a^2,\;\;
t^{[3]}_6=\frac{4 a^2}{1-b^2}
\eea
 for solution 2
($y_0(y_1,y_2)=y_1|y_2|$). Note that for both solutions, if $b<1$, all the $t_i^{[2,3]}$'s
are positive.

We then obtain 
\begin{align}
F^{(1)}_6(0)
&=\ln2\ln(1-b)
-2\ln2 \ln(1+b)\notag\\
&-2\ln(1-b)\ln(1+b)
+\frac{1}{2}(\ln(1-b))^2
+3(\ln(1+b))^2,\label{F6_1}
\end{align}
for solution 1 and
\begin{align}
F^{(1)}_6(0)
&=\frac{3}{2}\left\{(\ln(1-b))^2+(\ln(1+b))^2\right\}
-2\ln(1-b)\ln(1+b),\label{F6_2}
\end{align}
for solution 2.

The divergent piece of the amplitude becomes 
\bea
&&-\frac{\sqrt{\lambda}}{2\pi}\frac{6}{\epsilon^2}\left(\frac{\mu}{2a}\right)^{\epsilon}
((1+\frac{\epsilon}{2}(1-\ln 2))(1+\frac{\epsilon}{6}\ln (1-b)(1+b)^3)
\nonumber\\&&
+\frac{\epsilon^2}{4}(\ln^2(1+b)+\frac{1}{6}\ln^2(1-b))+\frac{\epsilon^2}{12}\ln 2(\frac{\ln 2}{
2}-\ln(1+b)))
\eea
for solution 1 and 
\bea
&&-\frac{\sqrt{\lambda}}{2\pi}\frac{6}{\epsilon^2}\left(\frac{\mu}{2a}\right)^{\epsilon}
((1+\frac{\epsilon}{2}(1-\ln 2))(1+\frac{\epsilon}{3}\ln (1-b^2))\nonumber\\&&
+\frac{\epsilon^2}{8}(\ln^2(1+b)+\ln^2(1-b)))\label{fieldtheory}
\eea
for solution 2. The finite remainder part can be rewritten 
as
\be
-\frac{\sqrt{\lambda}}{2\pi}\frac{6}{\epsilon^2}\left(\frac{\mu}{2a}\right)^{\epsilon}
(-\frac{\epsilon^2}{12}F_6^{(1)}(0))
\label{remainder}
\ee

Then, reintroducing the general dependence of $\lambda$ at finite coupling, we can 
write the total result for these 6 point amplitudes as 
\bea
&&{\cal M}_6=\frac{{\cal A}_6}{{\cal A}_{6,tree}}=d(\lambda)\exp(-\frac{3}{4\epsilon^2}
f^{-2}(\lambda(\frac{\mu}{2a})^{2\epsilon}))\exp(-\frac{3}{2}g^{-1}(\lambda(\frac{\mu}{2a})^{2
\epsilon}))\nonumber\\
&& \times
\left(\frac{1}{b+1}\right)^{\frac{3}{2}g(\lambda)+\frac{3}{4}\frac{f^{-1}(\lambda)}{\epsilon}
+\frac{3}{2}f(\lambda)\ln\frac{\mu}{2a}+\frac{f(\lambda)}{4}\ln 2}
\left(\frac{1}{1-b}\right)^{\frac{g(\lambda)}{2}+\frac{f^{-1}(\lambda)}{4\epsilon}+
\frac{f(\lambda)}{4}\ln\frac{\mu^2}{4a^2}\frac{(1+b)^2}{2}}
\eea
for solution 1 and
\bea
&&{\cal M}_6=\frac{{\cal A}_6}{{\cal A}_{6,tree}}=d(\lambda)\exp(-\frac{3}{4\epsilon^2}
f^{-2}(\lambda(\frac{\mu}{2a})^{2\epsilon}))\exp(-\frac{3}{2}g^{-1}(\lambda(\frac{\mu}{2a})^{2
\epsilon}))\nonumber\\
&& \times
\left(\frac{1}{b+1}\right)^{g(\lambda)+\frac{f^{-1}(\lambda)}{2\epsilon}+f(\lambda)\ln
\frac{\mu}{2a}}\left(\frac{1}{1-b}\right)^{g(\lambda)+\frac{f^{-1}(\lambda)}{2\epsilon}
+\frac{f(\lambda)}{2}\ln \frac{\mu^2}{4a^2}(1+b)}
\eea
for solution 2. Here $d(\lambda)$ contains finite constant factors.
The first line is equal in the two expressions, and is a IR divergent piece 
depending on the overall scale of the momentum. The second line can be rewritten as 
\be
\left(\frac{t_2^{[2]}}{2t_3^{[2]}}\right)^
{\frac{3}{4}g(\lambda)+\frac{3}{8}\frac{f^{-1}(\lambda)}{\epsilon}
+\frac{3}{4}f(\lambda)\ln\frac{\mu}{2a}+\frac{f(\lambda)}{8}\ln 2}\left(\frac{t_1^{[2]}}{
2t_3^{[2]}}\right)^{\frac{g(\lambda)}{2}+\frac{f^{-1}(\lambda)}{4\epsilon}+
\frac{f(\lambda)}{4}\ln\frac{\mu^2}{2t_2^{[2]}}}
\ee
for solution 1 and 
\be
\left(\frac{t_2^{[2]}}{t_2^{[3]}}\right)^{g(\lambda)+\frac{f^{-1}(\lambda)}{2\epsilon}+f(\lambda)\ln\frac{\mu}{2a}}\left(\frac{t_3^{[2]}}{t_2^{[3]}}\right)^{
g(\lambda)+\frac{f^{-1}(\lambda)}{2\epsilon}
+\frac{f(\lambda)}{2}\ln \frac{\mu^2}{t_2^{[2]}}}
\ee
for solution 2. This rewriting is similar to the one performed for the 4 point function 
in \cite{ns}, and as there, it relies on the nontrivial 
cancellation of the leading $\ln^2$ terms between the divergent part and the finite remainder
(in this case, $\ln^2(1-b)$ and $\ln^2(1+b)$ terms), without which one could not rewrite
the amplitude as this power law. 

We can then take the limit $b\rightarrow 1$ ($a$ is fixed), which takes several of the 
$t_i^{[r]}$ parameters to infinity, similar to the $s$ fixed, $t\rightarrow -\infty, u
\rightarrow +\infty$ limit taken by \cite{ns}, and since the tree amplitude 
${\cal A}_{6,tree}$ also 
behaves like a power law, we also get a Regge-like behaviour of the 6 point amplitude
${\cal A}_6$, $\sim (t_i^{[2]})^{\alpha(t_2^{[2]})}$, where
$t_i^{[2]}$ is a parameter that goes to infinity and $t_2^{[2]}$ stays finite. The physical significance of this result is not clear, but since this power law behaviour doesn't seem to 
hold for an arbitrary high energy limit (some of the t parameters becoming 
infinite, others staying finite, and for arbitrary values), 
it seems to suggest that the cases treated here have a 
Regge-like explanation as for the 4-point function, in terms of an exchanged particle.

We also observe that if $b<1$, since all the $t_i^{[2,3]}$'s are positive, the amplitude
is real, whereas if $b>1$ the amplitude becomes complex.

So we have a mismatch between the AdS and field theory results. But all the 
solutions that we wrote were obtained by cutting and gluing, so there is a potential 
problem on the line on which we glue. We will try to understand the $y_0=y_1|y_2|$
solution, since it is easiest, and the mismatch is smallest. 

There could be potential delta functions, $\delta(y_2)$, in the equation of motion 
in static gauge, coming from $\partial_2^2y_0$. Other than these potential terms, the 
equations of motion are the same for our solution as for the 4 point function solution, 
thus are satisfied (since the solutions were obtained by using symmetries of the action).

Since the terms are of the type $\delta(y_2)$, anything multiplied 
by $y_2$ gives zero, thus we can put $y_2=0$ after taking derivatives. We only look for 
$\partial_2^2 y_0$ terms, the only ones that give the delta functions. We also substitute
$\partial_1 y_0=0, \partial_2r=0$ after taking derivatives, 
since both are proportional to $y_2$. 

Then potential delta function terms in the r equation of motion coming from (\ref{action}) 
are contained in 
\be
\partial_2[\frac{\partial_2 r+\partial_1y_0 (\partial_1 r \partial_2 y_0-\partial_2 r
\partial_1 y_0)}{r^2\sqrt{1+(\partial_i r)^2-(\partial_i y_0)^2 -
(\partial_1r\partial_2 y_0-\partial_2 r\partial_1 y_0)^2}}]
\ee
but as we can easily see, after taking derivatives, keeping only $\partial_2^2y_0$ terms 
and substituting $y_2=0$ as above, we actually get zero. So there are no delta function 
terms in the r equation of motion. 

The $y_0$ equation of motion is 
\bea
&&-\frac{1}{r^2}
\partial_2[\frac{-2\partial_2 y_0-2\partial_1r (\partial_1 r \partial_2 y_0-\partial_2 r
\partial_1 y_0)}{2r^2\sqrt{1+(\partial_i r)^2-(\partial_i y_0)^2 -
(\partial_1r\partial_2 y_0-\partial_2 r\partial_1 y_0)^2}}]\nonumber\\&&
-\frac{1}{r^2}
\partial_1[\frac{-2\partial_1 y_0+2\partial_2r (\partial_1 r \partial_2 y_0-\partial_2 r
\partial_1 y_0)}{2r^2\sqrt{1+(\partial_i r)^2-(\partial_i y_0)^2 -
(\partial_1r\partial_2 y_0-\partial_2 r\partial_1 y_0)^2}}]=0
\eea
and again keeping only $\partial_2^2y_0$ terms and putting $y_2=0$ after taking the derivatives
we get, after a bit of algebra, the source (boundary) term
\be
\frac{1}{r^2}\partial_2[\frac{\partial_2 y_0+(\partial_1 r)^2\partial_2 y_0}
{\sqrt{1+(\partial_i r)^2-(\partial_i y_0)^2 -
(\partial_1r\partial_2 y_0-\partial_2 r\partial_1 y_0)^2}}]
=\frac{y_1\delta(y_2)}{(1-y_1^2)^3}
\ee

So one needs to add a source term in the $y_0$ equation of motion (but not in the
r equation of motion) that cancels this term. In other words, we have an extra boundary 
condition at $y_2=0$, a boundary condition in $y_0$, but not in $r$. The boundary condition is 
that $y_0(y_2=0)=0$ (but $r$ is arbitrary).

\section{Mismatch interpretation}

It is easy to see what would be the interpretation of the boundary condition
identified in the previous section. The external 
boundary of the Wilson loop (for which $r$=0) is mapped by T duality to physical (on-shell)
external momenta of the amplitude. T duality will map the line $y_0=y_2=0$, $\Delta y_1
=2$ to a momentum $k^{\mu}$: $(E=0,p^2=0$, $p^1=2)$, 
which is therefore virtual, being spacelike. 
Moreover, the line $y_2=0$ has varying $r$, which is equal to zero only at the ends. Therefore
this momentum is not external (external momenta are defined on the $r$=0 boundary). 

\begin{figure}[ht]
\begin{center}
\resizebox{120mm}{!}{\includegraphics{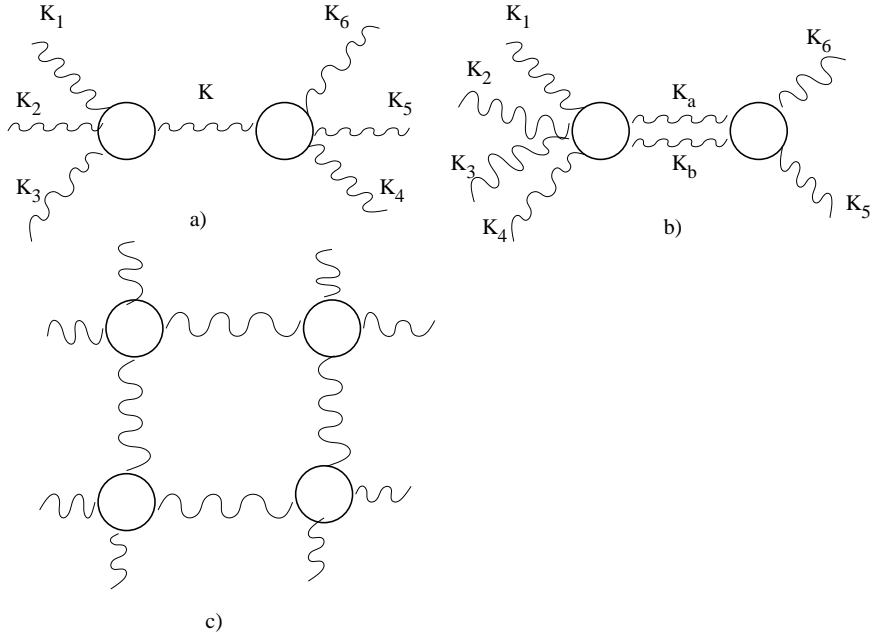}}
\end{center}
\caption{a) Conjectured amplitude calculated by solution 2. b) Conjectured amplitude 
calculated by solution 1. c) Conjectured amplitude calculated by 8-point function 
solution.}\label{amplit}
\end{figure}

Thus we propose the interpretation that the AdS amplitude we calculated actually 
corresponds to the following field theory amplitude. Amplitude for three external lines
to go into the virtual line $k^{\mu}$: $(E=0,p^2=0$, $p^1=2)$,
 followed by amplitude for this virtual
line to go into other three external lines, as in Fig.\ref{amplit}a.

It could however also be that there simply is a mismatch between the BDS formula and 
the dual prescription. Indeed, recently \cite{am3} found a mismatch for the ${\cal M}_n$
amplitude at large $n$. They also suggested that since 4- and 5-point amplitudes are 
determined by conformal symmetry \cite{dhks,am3}, there could in principle be dissagreements starting at the 6-point
amplitude.

One could ask whether the mismatch between the first line in (\ref{adsresult}) and 
(\ref{fieldtheory}) plus (\ref{remainder}) can be fixed. At first sight this seems 
encouraging. Indeed, \cite{buch} showed that the divergent terms in the BDS formula 
can be obtained from the contribution near the cusps (corners) of the Wilson loop. 

The four corners of the Alday-Maldacena solution have thus the correct behaviour, and 
they are the same for us, so they are guaranteed to match. But the two extra cusps
on the $y_0=y_2=0$ line are potentially problematic. So we could ask whether it is 
enough to subtract the contribution at our (unsatisfactory) cusps and add the 
correct cusp behaviour. 

The correct cusp behaviour is, according to \cite{buch}
\bea
&&\sum_{i=1,4}(-\frac{\sqrt{\lambda}}{2\pi})
\frac{1}{\epsilon^2}C(\epsilon)\left(\frac{\mu^2}{s_{i,i+1}}\right)^{\epsilon/2}
\nonumber\\
&&= (-\frac{\sqrt{\lambda}}{2\pi})\left(\frac{\mu}{2a}\right)^{\epsilon}
[\frac{2}{\epsilon^2}+\frac{1-\ln 2}{\epsilon}
+\frac{1}{2}\ln^2(1-b)+\frac{1}{2}\ln^2(1+b)\nonumber\\&&
+\frac{1}{\epsilon}(\ln(1+b)+\ln(1-b))
+\frac{1-\ln 2}{2}(\ln (1+b)+\ln (1-b))]
\eea
and we see that at least the b-independent, epsilon-divergent terms are the ones needed 
for the mismatch. 

The contribution of the fake cusps is evaluated in the Appendix. 
It is found to be of order $1/\epsilon$ as needed (since we are missing the $1/\epsilon^2$
term), but the $b$ dependence is not the one we wanted.
 That means that unfortunately, 
the missing contribution is not localized at the two fake cusps only. 

We then go back to the interpretation of the AdS amplitude as field theory amplitude with 
an intermediate virtual line and try to understand it better.

If the amplitude we are calculating involves one intermediate virtual line, that means  
that in order to complete the full 6-point amplitude we are missing amplitudes where the
intermediate virtual line is replaced by 2, 3, ... (any number $>1$) of intermediate virtual 
lines. 

The separation of the total 6-point amplitude in amplitudes with any number of intermediate
lines is familiar from the optical theorem. The optical theorem is a diagramatic equality
based on the operatorial relation $-i(T-T^\dag)=T^\dag T$, where $T=(S-1)/i$ is the T matrix. 
The optical theorem states that (twice) the imaginary part of the 6 point amplitude is 
equal to the sum of the cut amplitudes with 1,2,3,... (any number of) intermediate lines,
where cut means putting the lines on-shell, i.e. replacing (for a scalar propagator)
\be
\frac{1}{p^2-m^2+i\epsilon}\rightarrow -2\pi i \delta(p^2-m^2)
\ee
In order to have such a contribution, we need to have at least 
an integration over a (loop) momentum for the intermediate lines, which means that the 
1 particle cut never contributes. 

So the imaginary part of the 6-point amplitude is given by the sum over 2,3,... particle 
cuts. But the BDS formula states that the 6-point amplitude is real if we have 
all $t_i^{[2,3]}$'s positive, as is the case for us if $b<1$, therefore
the sum of the 2,3,... particle cuts in our case if $b<1$ must be zero. 

But the contribution we are missing is one where in the same diagrams
we don't cut the propagators, but we 
compute the whole integral, thus can be potentially nonzero.

Let us also note that we can interpret in a similar manner the 6 point amplitude 
corresponding to the $y_0(y_1,y_2)=1/2(|y_1y_2|+y_1y_2-|y_1|y_2+y_1|y_2|)$
solution. In a similar way, we see that it has two extra boundaries at $y_1=0=y_0, y_2<0$
and $y_2=0=y_0,y_1>0$, corresponding to 2 spacelike (virtual) momenta $k_a=(0,0,-1)$ and
$k_b=(0,1,0)$. Therefore this time the conjectured corresponding field theory amplitude 
is the amplitude for 4 external lines to go into the two virtual momenta $k_a$ and $k_b$, 
followed by the amplitude for the two virtual momenta to go into other two external 
lines, as in Fig.\ref{amplit}b).

Therefore we conjecture that any extra boundary condition for the Wilson loop, 
defined on a line, that fixes the 
$y_\mu$'s ($\mu=0,1,2,3$), but not $r$, corresponds to an intermediate virtual momentum line 
with $k^{\mu}=\Delta y^{\mu}/(2\pi)$. A priori one can have a Wilson loop with many such 
boundary conditions, and therefore get an amplitude with many intermediate virtual lines, 
but then the AdS calculation is probably less useful (it is less useful to know only a very 
particular set of Feynman diagrams). That is why we have focused on the solution with a 
single intermediate momentum line.

\section{Eight point function}

\begin{figure}[bthp]
\begin{center}\includegraphics{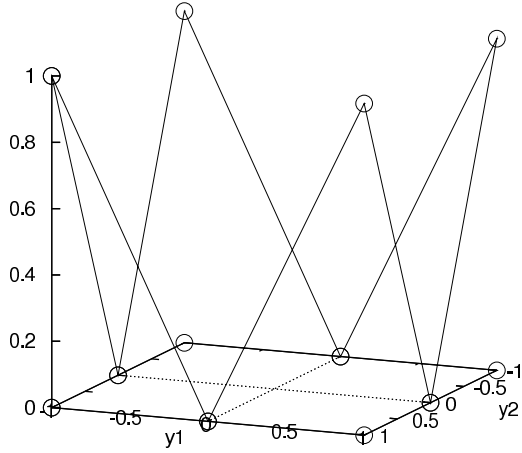}\end{center}
\caption{Configuration for 8-point amplitude solution, $y_0=|y_1y_2|$}\label{8pt}
\end{figure}

\begin{figure}
\begin{minipage}{7cm}\vspace{-.7cm}\includegraphics{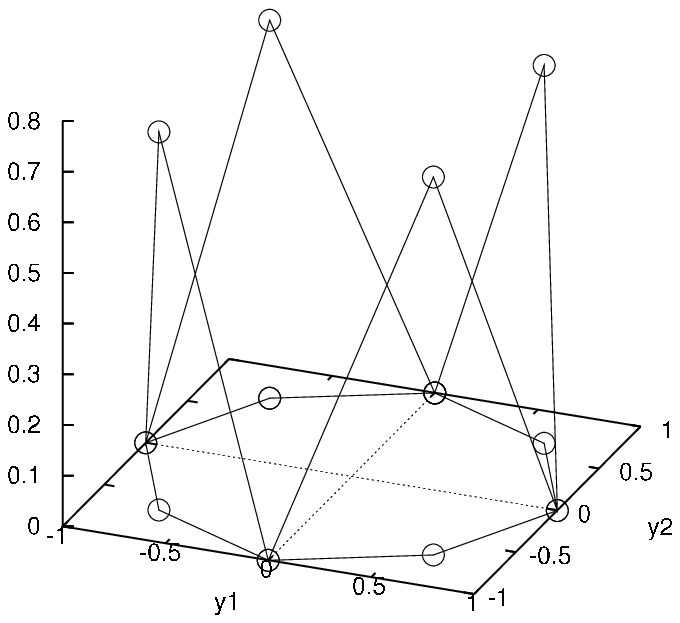}\end{minipage}
\begin{minipage}{5cm}\includegraphics{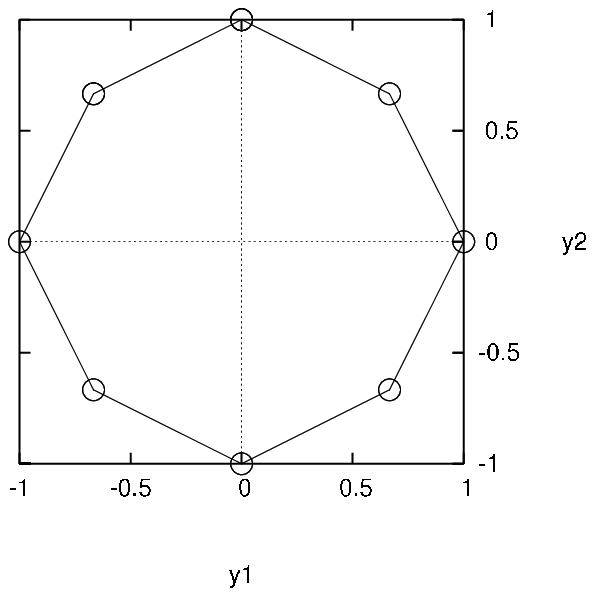}\end{minipage}
\caption{Configuration after the Lorentz boost
in the 04 plane for the solution $y_0=|y_1y_2|$, with a=1, b=0.5.}
\label{8ptAL}
\end{figure}

We can generate also an 8-point amplitude from the Alday-Maldacena solution in a  
manner similar to the 6-point functions. The solution is
\be
y_0(y_1,y_2)=|y_1y_2|;\;\;\;
r(y_1,y_2)=\sqrt{(1-y_1^2)(1-y_2^2)}
\ee
which is depicted in Figure \ref{8pt}.
We can again Lorentz boost the solution in the $Y_4$ embedding 
coordinate of AdS and obtain the solution drawn in Figure \ref{8ptAL}. 
It is the same solution as in (\ref{boosted}), except now the $\pm$ is $=sgn(y_1y_2)$. 
From it we can derive the external momenta
\bea
&&k_{1}=\left(-\frac{a b}{b+1},-\frac{a}{b+1},-\frac{a\sqrt{b^2+1}}{b+1}\right),\;\;\;
k_{2}=\left(\frac{a b}{b+1},-\frac{a}{b+1},\frac{a \sqrt{b^2+1}}{b+1}\right),\nonumber\\
&&k_{3}=\left(\frac{a}{b+1},-\frac{a b}{b+1},-\frac{a\sqrt{b^2+1}}{b+1}\right),\;\;\;
k_{4}=\left(\frac{a}{b+1},\frac{a b}{b+1},\frac{a\sqrt{b^2+1}}{b+1}\right),\nonumber\\
&&k_{5}=\left(\frac{a b}{b+1},\frac{a}{b+1},-\frac{a\sqrt{b^2+1}}{b+1}\right),\;\;\;
k_{6}=\left(-\frac{a b}{b+1},\frac{a}{b+1},\frac{a\sqrt{b^2+1}}{b+1}\right),\nonumber\\
&&k_{7}=\left(-\frac{a}{b+1},\frac{a b}{b+1},-\frac{a\sqrt{b^2+1}}{b+1}\right),\;\;\;
k_{8}=\left(-\frac{a}{b+1},-\frac{a b}{b+1},\frac{a\sqrt{b^2+1}}{b+1}\right).
\eea
The momentum invariants are then
\be
t^{[2]}_{\text{odd}}=\frac{4 a^2}{(b+1)^2},\;\;\;
t^{[2]}_{\text{even}}=2 a^2,\;\;\;
t^{[3]}_{i}=\frac{4 a^2}{b+1},\;\;\;
t^{[4]}_{\text{odd}}=\frac{8 a^2}{(b+1)^2},\;\;\;
t^{[4]}_{\text{even}}=4 a^2
\ee

With these values, the finite remainder function is 
\be 
F_8^{(1)}(0)=4 \ln^2(b+1)-4\ln 2 \ln (b+1) -\frac{\pi^2}{6}
\ee
where we have used the relation
\be
\text{Li}_2(1/2)=\frac{\pi^2}{12}-\frac{1}{2}(\ln 2)^2.
\ee

The divergent part of the amplitude, from (\ref{divamp}) is found to be
\bea
&&\ln {\cal M}_{n,div}=-\frac{4\sqrt{\lambda}}{\pi \epsilon^2}
\left(\frac{\mu}{2a}\right)^\epsilon((1+\frac{\epsilon}{2}(1-\ln 2))(1+\frac{\epsilon}{2}\ln
(b+1)+\frac{\epsilon\ln 2}{4})\nonumber\\&&
+\frac{\epsilon^2}{16}(4\ln^2(b+1)+\ln^22))
\eea

On the other hand, the AdS result is found to be obtained by multiplying the second term 
in (\ref{result2}) by a factor of 2, thus the final result in (\ref{adsresult}), with 
the last line multiplied by a factor of 2, i.e.
\bea
&& -\frac{\sqrt{\lambda}}{2\pi}(2\pi^2\frac{\mu^2}{4a^2})^{\epsilon/2}\left[\frac{4}{\epsilon
^2}+\frac{2}{\epsilon}\ln (1-b^2)+\frac{2}{\epsilon}(1-\ln 2)\right.\nonumber\\&&\left.
+(1-\ln 2)\ln (1-b^2)
+2\ln (1-b)\ln (1+b)\right.\nonumber\\&&
\left.+\frac{2}{\epsilon}\ln\frac{1+b}{1-b}+2\frac{1+\ln 2}{2}\ln \frac
{1+b}{1-b}-2{\rm Li}_2\left(\frac{1-b}{2}\right)+2{\rm Li}_2\left(\frac{1+b}{2}\right)\right]
\eea
As evaluated in the appendix, the contribution of the subleading terms give also twice the 
subleading terms of solution 1 for the 6-point function, thus we have an extra $-2b$ in 
the square brackets.

The mismatch now is most dramatic, but again the explanation is that we have now 4 extra 
boundaries on which $y_0=0$, namely $y_1=0, y_2>0$; $y_1=0,y_2<0$; $y_2=0,y_1>0$; $y_2=0,
y_1<0$. They will correspond to 4 internal spacelike (virtual) momenta, thus giving the 
amplitude in Fig.\ref{amplit}c).

\section{Collinear limits}

The Alday-Maldacena solution can also be reinterpreted as the collinear limit of a higher $n$
point amplitude. That is, if we interpret the 4 sides of the Wilson loop not 
as a single external momentum, but as sets of momenta:
\be
k_1\rightarrow \sum_i k_i^{(1)};\;\;\;
k_2\rightarrow \sum_i k_i^{(2)};\;\;\;
k_3\rightarrow \sum_i k_i^{(3)};\;\;\;
k_4\rightarrow \sum_i k_i^{(4)}
\ee
and replace in $s=(k_1+k_2)^2$ and $t=(k_2+k_3)^2$. The result for such an n-point amplitude
is the Alday-Maldacena result as a function of $s$ and $t$, now defined as a function of 
the n momenta $k_i^{(a)}, a=1,2,3,4$.  This is a prediction of the AdS calculation, and we 
should check that it is indeed obtained from the BDS conjecture. We will see however that 
there are subtleties related to how we take the limit. 

We will now check that the BDS formula for the n-point functions
reproduces the 4 point result in the above collinear limit. 

Specifically, let us consider the case of 5 point amplitude and take $k_4=z k_P$ and 
$k_5=(1-z) k_P$, so that $k_P=k_4+k_5$ and take the limit $k_P^2\rightarrow 0$. This 
is a usual collinear limit. However, we already see that this is not quite how the limit
is taken in string theory. In the AdS computation, we have amplitudes that are already 
on-shell ($k_4^2=k_5^2=0$), and it is only $2k_4\cdot k_5=(k_4+k_5)^2$ that goes to zero.

For the 5 point function we have
\be
g_{5,i}=L_{5,i}=-\frac{1}{2}\ln\left(\frac{t_i^{[2]}}{t_{i+3}^{[2]}}\right)\ln\left(
\frac{t_{i+1}^{[2]}}{t_{i+2}^{[2]}}\right)
\ee
and ignoring subleading terms in $k_P^2$, we have the variables
\be
t_1^{[2]}=s_{1,2};\;\;\; t_2^{[2]}=s_{2,3};\;\;\;t_3^{[2]}=s_{3,4}=z s_{3,P};\;\;\;
t_4^{[2]}=s_{4,5}=k_P^2;\;\;\; t_5^{[2]}=s_{5,1}=(1-z)s_{P,1}
\ee
The momenta $(k_1,k_2,k_3,k_P)$ characterize the 4 point amplitude, with variables
\be
s_{1,2}=s_{3,P}=-s;\;\;\; s_{1,P}=s_{2,3}=-t
\ee
Then the finite remainder of the 5 point amplitude is
\bea
&&\frac{f(\lambda)}{4}F_5^{(1)}(0)=\frac{f(\lambda)}{8}\sum_{i=1}^5 L_{5,i}=
\frac{f(\lambda)}{8}[\ln^2\frac{s}{t}\nonumber\\
&&+\ln \frac{-s}{\mu^2}\ln z +\ln\frac{-t}{\mu^2}\ln (1-z)-\ln \frac{k_P^2}{\mu^2}
\ln z(1-z)+\ln z\ln (1-z)]
\eea
where we have introduced an arbitrary scale $\mu$ that we want to identify with the IR 
scale, in order to isolate the finite remainder of the 4 point function, the first term 
in the last equality. 

The divergent piece of the (log of the) 5 point amplitude is 
\bea
&&-\frac{5 f^{-2}(\lambda)}{8\epsilon^2}-\frac{5g^{-1}(\lambda)}{4\epsilon}
-(\frac{f^{-1}(\lambda)}{8\epsilon}+\frac{g(\lambda)}{4})[2\ln\frac{\mu^2}{s}+2\ln\frac{\mu^2}
{t}-\ln \frac{k_P^2z(1-z)}{\mu^2}]\nonumber\\
&&-\frac{f(\lambda)}{16}[2\ln^2\frac{\mu^2}{s}+2\ln^2\frac{\mu^2}{t}+\ln^2(\frac{k_P^2 z(1-z)}{
\mu^2})-2\ln z \ln (1-z)\nonumber\\&&
+2\ln \frac{s}{\mu^2}\ln z+ 2\ln\frac{t}{\mu^2}\ln (1-z)
-2\ln \frac{k_P^2}{\mu^2}\ln z(1-z)]
\eea
Adding up the 2 contributions the last line in the divergent piece cancels 
against the finite remainder and we get the 4 point 
amplitude with some extra terms
\bea
&&\ln {\cal M}_5\rightarrow \ln {\cal M}_4-\frac{f^{-2}(\lambda)}{8\epsilon^2}-
\frac{g^{-1}(\lambda)}{4\epsilon}
+\frac{f(\lambda)}{4}\ln z\ln (1-z)\nonumber\\&&
-\frac{f(\lambda)}{16}\ln^2\frac{k_4\cdot k_5}{\mu^2}+
(\frac{f^{-1}(\lambda)}{8\epsilon}+\frac{g(\lambda)}{4})\ln\frac{k_4\cdot k_5}{\mu^2}
\label{extra}
\eea

This computation agrees with the one loop result in \cite{bddk} since at one loop $g(\lambda
)=0$ and the extra terms are then
\be
\frac{f(\lambda)}{8}2\ln z\ln (1-z)-\frac{1}{8\epsilon^2}f^{-2}
\left(\lambda\left(\frac{\mu^2}{k_4
\cdot k_5}\right)^\epsilon\right)
\ee

However the extra terms are unfortunate from the point of view of the AdS calculation. 
The second line in (\ref{extra}) dissappears if we take $k_4\cdot k_5=\mu^2$, which is 
consistent, since both quantities go to zero. The $2\ln z\ln (1-z)$ can be rewritten as 
$1/2\ln k_4^2/k_P^2\ln k_5^2/k_P^2$ and thus is seen to be due to the fact that $k_4^2$ 
and $k_5^2$ were not zero from the begining, as was the case in the AdS computation. 
We are still left with the constant terms $-f^{-2}(\lambda)/8\epsilon^2-g^{-1}(\lambda)/
4\epsilon$ which arise from the corner of the AdS Wilson loop
and thus should dissappear if the collinear 
limit of the AdS calculation is done correctly (and 
before taking $\epsilon$ to zero). 

Next, we consider the 6 point amplitude and take the double collinear limit, $k_1=wk_Q$,
$k_2=(1-w) k_Q$, and $k_5=zk_P$ and $k_6=(1-z)k_P$. As before, $k_P^2$ and $k_Q^2$ are not
zero, but rather go to zero in the collinear limit. In this limit we obtain (dropping 
subleading $k_Q^2$ and $k_P^2$ terms
\bea
&& t_1^{[2]}=s_{1,2}=k_Q^2;\;\;\;t_2^{[2]}=s_{2,3}=(1-w) s_{Q,3};\;\;\;
t_3^{[2]}=s_{3,4}\nonumber\\
&& t_4^{[2]}=s_{4,5}=zs_{4,P};\;\;\; t_5^{[2]}=s_{5,6}=k_P^2;\;\;\;
t_6^{[2]}=s_{6,1}=(1-z)s_{P,1}=(1-z)ws_{P,Q}\nonumber\\
&& t_1^{[3]}=s_{4,P};\;\;\; t_2^{[3]}=s_{P,1}=ws_{P,Q};\;\;\;
t_3^{[3]}=(1-z)s_{P,Q}=(1-z)s_{3,4};\;\;\; t_i^{[3]}=t_{i+3}^{[3]}
\eea
The momenta $(k_P,k_Q,k_3,k_4)$ characterize the 4 point amplitude, with variables
\be
s_{P,Q}=s_{3,4}=s;\;\;\; s_{Q,3}=s_{P,4}=t
\ee
in the limit that $k_P$ and $k_Q$ are on-shell. Then the finite remainder term is 
\bea
&& \frac{f(\lambda)}{4}F_6^{(1)}(0)=\frac{f(\lambda)}{8}\sum_{i=1}^6 g_{6,i}=
\frac{f(\lambda)}{8}[\ln^2 \frac{s}{t}+\nonumber\\
&&+\ln\frac{s}{\mu^2}\ln w(1-z)+\ln\frac{t}{\mu^2}\ln z(1-w)
-\ln\frac{k_P^2}{\mu^2}\ln z(1-z)-\ln\frac{k_Q^2}{\mu^2}\ln w(1-w)\nonumber\\
&&+\ln z\ln (1-z)+\ln w\ln (1-w)+\ln w\ln (1-z)]
\eea
and the divergent part of the (log of the) 6-point amplitude is
\bea
&&-\frac{3f^{-2}(\lambda)}{4\epsilon^2}-\frac{3g^{-1}(\lambda)}{2\epsilon^2}
-(\frac{f^{-1}(\lambda)}{8\epsilon}+\frac{g(\lambda)}{4})(2\ln\frac{\mu^2}{s}+2\ln
\frac{\mu^2}{t}-\ln\frac{k_Q^2}{\mu^2}w(1-w)\frac{k_P^2}{\mu^2}z(1-z))\nonumber\\
&&-\frac{f(\lambda)}{16}[2\ln^2\frac{\mu^2}{s}+2\ln^2\frac{\mu^2}{t}+
\ln^2\frac{k_P^2}{\mu^2}z(1-z)\nonumber\\&&
+\ln^2\frac{k_Q^2}{\mu^2}w(1-w)-2\ln z\ln (1-z)-2\ln w
\ln (1-w)\nonumber\\
&&+2\ln w\ln (1-z)+2\ln\frac{t}{\mu^2}\ln z(1-w)+2\ln \frac{s}{\mu^2}\ln w(1-z)\nonumber\\&&
-2\ln\frac{k_P^2}{\mu^2}\ln z(1-z)-2\ln\frac{k_Q^2}{\mu^2}\ln w(1-w)]
\eea
Adding the two contributions the last two lines of the divergent part cancel against 
terms in the finite remainder and we get
\bea
&&\ln {\cal M}_6\rightarrow \ln {\cal M}_4\nonumber\\&&
+\frac{f(\lambda)}{8}2\ln w\ln (1-w)
-\frac{1}{8\epsilon^2}f^{-2}\left(\lambda\left(\frac{\mu^2}{k_1\cdot k_2}\right)^{\epsilon}
\right)-\frac{1}{4\epsilon}g^{-1}\left(\lambda\left(\frac{\mu^2}{k_1\cdot k_2}\right)^{\epsilon}
\right)\nonumber\\
&&+\frac{f(\lambda)}{8}2\ln z\ln (1-z)
-\frac{1}{8\epsilon^2}f^{-2}\left(\lambda\left(\frac{\mu^2}{k_5\cdot k_6}\right)^{\epsilon}
\right)-\frac{1}{4\epsilon}g^{-1}\left(\lambda\left(\frac{\mu^2}{k_5\cdot k_6}\right)^{\epsilon}
\right)
\eea
i.e., the sum of the contributions of the two collinearities, as expected.

\section{Conclusions}

In this paper we have analyzed 6 point amplitudes for gluon scattering at strong coupling 
and large N in ${\cal N}=4$ SYM, using AdS-CFT, following the prescription of \cite{am}. 
We have used the BDS conjecture together with the 
strong coupling value of the functions $f(\lambda)$ and $g(\lambda)$ calculated in 
\cite{am} to predict what the AdS results should give. For the AdS calculation,  we have 
analyzed solutions obtained by symmetries, cutting and gluing. We have obtained a 
mismatch, due to the fact that the AdS solutions contain extra boundary conditions. 

The boundary conditions are that $y_0=0$ on an internal line where $r$ is not fixed, and 
we have interpreted them as having a fixed intermediate virtual momentum line in the 
amplitude. Thus we propose that
the AdS computation calculates only a certain part of the 6-point amplitudes.
It would be interesting if one could calculate the gauge theory value for the 
corresponding amplitude, in order to really test our proposal. 

It could also be that there is an actual dissagreement between the BDS conjecture 
and the dual computation. In \cite{am3} it was suggested that a dissagreement could start 
at $n$-point amplitudes with $n\geq 6$. The 4- and 5-point amplitudes are fixed by 
conformal invariance \cite{dhks,am3}, but a dissagreement was found at $n\rightarrow\infty$. 

The 6-point functions analyzed here do not cover the general external momenta (we have 
only 2 variables, instead of 8), and in particular we found that for these momenta we 
obtain a kind of Regge behaviour, where if we take some of the $t_i$'s to infinity by 
taking $b\rightarrow 1$ (which keeps the rest of the $t_j$'s fixed) we have 
${\cal A}\sim (t_i)^{\alpha (t_j)}$. It would be interesting to understand the physical 
significance of this result. 

We have also treated an 8-point function for completeness, which can be obtained similarly.
In this case however, the mismatch is more dramatic, which we understood from our conjectured
picture for the extra boundary conditions: the gauge theory amplitude contains only a small 
part of the possible Feyman diagrams. 

The calculation of \cite{am} can be reinterpreted as being a higher n-point amplitude, where
the momenta are collinear, such that they form four groups. This implies that there should 
be a way to take the collinear limit that should avoid extra terms. We have calculated the 
natural collinear limit of the 5- and 6-point BDS amplitudes, and we have found that we can 
get rid of most, but not all the extra terms. The issue needs therefore to be understood 
further, but this can only be done if we have a solution with correct extra cusps (for our
solutions, as we saw, the extra cusps did not have the right BDS behaviour). 

{\bf Acknowledgements} H.N. would like to thank Radu Roiban and Gabriele Travaglini,
 and D.A. would like to thank Niklas Beisert and Stefan Theisen  for  discussions.
This research  has been done with partial support from MEXT's program
"Promotion of Environmental Improvement for Independence of Young Researchers"
under the Special Coordination Funds for Promoting Science and Technology. D.A.
also acknowledges the hospitality of the Albert Einstein Institute in Potsdam during the last stages of 
this research, as well as support from NSERC of Canada.

\newpage

\section{Appendix}

\subsection{Integrals}

In this appendix we show how to compute integrals 
necessary for the AdS 6 point amplitudes. The calculation proceeds along the same line as the 
calculation of the similar integral in the appendix of \cite{am}. First we consider 
the integral relevant for the leading term (formally of order 1) in (\ref{regaction})
\be
I=\int_{-\infty}^{\infty}
du_1 du_2
(\cosh u_1 \cosh u_2 + \beta \sinh u_1 \sinh u_2)^\epsilon
\ee
and expanding in $\beta$ we get 
\be
\sum_{l=0}^{\infty}\int_{-\infty}^{+\infty}du_1\int_{-\infty}^{+\infty} du_2
\beta^l\frac{\Gamma(\epsilon+1)}{\Gamma(\epsilon+1-l)l!}(\cosh u_1 \cosh u_2)^{\epsilon}
(\tanh u_1 \tanh u_2)^l
\ee
We split the $u_1,u_2$ integrals into $(-\infty,0)$ and $(0,+\infty$) and use that $\beta
=\pm b$ is constant on those intervals. Then using
\be
\int_0^{+\infty} du (\cosh u)^\epsilon(\tanh u)^l=\frac{\Gamma(\frac{l+1}{2})\Gamma(-\frac{\epsilon}{2})}
{2\Gamma(\frac{1+l-\epsilon}{2})}
\ee
we get for solution 1 ($\pm=-$ if $u_1>0,u_2<0$ and $\pm=+$ otherwise)
\be
I=\sum_{l=0}\left(2(1+(-1)^l)+(1-(-1)^l)\right)\frac{\Gamma(\epsilon+1)}{\Gamma(\epsilon+1-
l)l!}b^l\left(\frac{\Gamma(\frac{l+1}{2})\Gamma(-\frac{\epsilon}{2})}{2\Gamma(\frac{1+l-
\epsilon}{2})}\right)^2
\label{sums}
\ee
and doing the sums we get 
\be
\frac{\pi \Gamma[-\frac{\epsilon}{2}]^2}{\Gamma[\frac{1-\epsilon}{2}]^2}
~_2F_1(\frac{1}{2},-\frac{\epsilon}{2},\frac{1-\epsilon}{2};b^2)+
\frac{2 b}{\epsilon}
~_3F_2(1,1,\frac{1-\epsilon}{2};\frac{3}{2},1-\frac{\epsilon}{2};b^2)
\ee
For this last step write the definitions of the hypergeometric functions as sums and 
then prove that the terms in the two expressions are the same. 

For solution 2, $\pm=+$ if $u_2>0$ and $\pm=-$ if $u_2<0$, we get only the first term in
(\ref{sums}), i.e. $2(1+(-1)^l)$, and not the $(1-(-1)^l)$ term, and consequently the ${}_3F_2$
term dissappears in the final result.

For the 8 point function solution, $\pm=+$ if $u_1u_2>0$ and $\pm=-$ if $u_1u_2<0$, and 
we get twice the $(1-(-1)^l)$ term, consequently twice the ${}_3F_2$ term in the final result.

A more general integral, needed for the calculation of the subleading terms is 
\bea
&&I=\int_0^{\infty} du_1du_2 (\cosh u_1 \cosh u_2 + b
 \sinh u_1 \sinh u_2)^a \times
\nonumber\\&& \times\cosh^m u_1
\cosh ^n u_2\tanh^p u_1 \tanh^q u_2=I_{even}+I_{odd}
\eea

Then 
\bea
&&
I_{even}=\frac{1}{4}B\left(\frac{p+1}{2},-\frac{a+m}{2}\right)
B\left(\frac{q+1}{2},-\frac{a+n}{2}\right)\nonumber\\
&&\times {}_4F_3\left(\{\frac{p+1}{2},\frac{q+1}{2},\frac{1-a}
{2}\};\{\frac{1}{2},\frac{p+1}{2}-\frac{a+m}{2},\frac{q+1}{2}
-\frac{a+n}{2}\};b^2\right)\nonumber\\
&&I_{odd}=\frac{ab}{4}\frac{\Gamma(-\frac{a+m}{2})\Gamma(-\frac{a+n}{2})}{\Gamma(\frac{2+p-a-m}{2})\Gamma(\frac{
2+q-a-n}{2})}\nonumber\\
&&\times {}_4F_3\left(\{\frac{p+2}{2},\frac{q+2}{2},1-\frac{a}
{2},\frac{1-a}{2}\};\{ \frac{3}{2},\frac{p+2-a-m}{2},\frac{
q+2-a-n}{2}\};b^2\right)
\eea
where $a+n,a+m<0,p+1,q+1>0$. For $m=n=p=q=0$ we get the previous
integral, and for $m=2$, $n=p=q=0$ we get 
\bea
&&I=\frac{1}{4\Gamma(\frac{1-a}{2})}\left\{\frac{\pi\Gamma(-1-
\frac{a}{2})\Gamma(-\frac{a}{2})}{\Gamma(-\frac{1+a}{2})}
{}_2F_1(\frac{1}{2},-\frac{a}{2},-\frac{1+a}{2};b^2)
\right.\nonumber\\&&\left.
+\frac{2^{2+a}(1+a)b\pi\Gamma(-2-a)}{\Gamma(\frac{3}{2})
\Gamma(-\frac{a}{2})}{}_3F_2(1,1,\frac{1-a}{2};\frac{3}{2},
-\frac{a}{2};b^2)\right\}
\eea

\subsection{Subleading terms in the action}

We write the terms in (\ref{regaction}) as
\be
-iS=B_{\epsilon}\int_{-\infty}^{+\infty}du_1du_2
\frac{1}{(r/a)^{\epsilon}}(1+\epsilon I_1+\epsilon^2 I_2+...)
=B_{\epsilon}\int_{-\infty}^{+\infty}du_1du_2 F_b(u_1,u_2)
\ee
thus the integrand splits as 
\be
F_b(u_1,u_2)=F_b^{(0)}+\epsilon F_b^{(1)}(u_1,u_2)+\epsilon^2
F_b^{(2)}(u_1,u_2)+...
\ee

For the 
solution 2, we can reduce the integration to integration 
from 0 to infinity by using the symmetries. We get
\be
-iS=B_{\epsilon}\int_0^{\infty}\int_0^{\infty}du_1du_2\left\{
F_b(u_1,u_2)+F_b(-u_1,u_2)+F_{-b}(u_1,-u_2)+F_{-b}(-u_1,u_2)
\right\}
\ee
but because $F_b(-u_1,u_2)=F_b(u_1,-u_2)=F_{-b}(u_1,u_2)$ 
we get
\be
-iS=B_{\epsilon}
\int_0^{\infty}\int_0^{\infty}du_1du_2\;\; 2 \left\{
F_b(u_1,u_2)+F_{-b}(u_1,u_2)\right\}
\ee
which is the same result as for the 4-point function. Thus, 
as is the case there, the subleading terms just give a $+1$
added inside the square brackets in (\ref{adsresult}).

For the solution 1, we have
\be
-iS=B_{\epsilon}
\int_0^{\infty}\int_0^{\infty}du_1du_2\left\{
F_{-b}(u_1,u_2)+F_b(-u_1,u_2)+F_{b}(u_1,-u_2)+F_{b}(-u_1,-u_2)
\right\}
\ee
and using the symmetries, we get
\bea
&&-iS=B_{\epsilon}
\int_0^{\infty}\int_0^{\infty}du_1du_2\left\{2
F_{b}(u_1,u_2)+2
F_{-b}(u_1,u_2)-(F_{b}(u_1,u_2)-F_{-b}(u_1,u_2))
\right\}\nonumber\\
&&=-iS^{4-point}-B_{\epsilon}
\int_0^{\infty}\int_0^{\infty}du_1du_2(F_b(u_1,u_2)-F_{-b}
(u_1,u_2))
\eea
Then the order $\epsilon$ term (from $F^{(1)}_b(u_1,u_2)$)
in the difference gives
\bea
&&-i \Delta S^{(1)}=-(b^2-1)\frac{2^{\epsilon}(\epsilon -1)
b\pi \Gamma(-\epsilon)}{4\Gamma (\frac{3-\epsilon}{2})\Gamma
(\frac{3}{2})\Gamma(-\frac{\epsilon-2}{2})}{}_3F_2(1,1,\frac{
3-\epsilon}{2};\frac{3}{2}-\frac{\epsilon-2}{2};b^2)\nonumber
\\&& +b^2\frac{b}{\epsilon -2}{}_3F_2(1,1,\frac{3-\epsilon}{2}
; \frac{3}{2},2-\frac{\epsilon}{2};b^2)
\eea
which in the limit of $\epsilon\rightarrow 0$ becomes $
b/\epsilon+...$. Then from the relation (\ref{regaction})
we can check that the order $\epsilon^2$ term in the action 
does not contribute (goes to zero). 

Thus for solution 1, the contribution of subleading terms adds
up to a $(-b)$ inside the square brackets in (\ref{adsresult}).

For the 8-point function, we get 
\bea
&& -iS=B_{\epsilon}
\int_0^{\infty}\int_0^{\infty}du_1du_2\left\{
2F_{b}(u_1,u_2)+2F_{-b}(u_1,u_2)+2F_{b}(u_1,-u_2)-2F_{-b}(u_1,u_2)\right\}\nonumber\\
&&=-iS^{4-pt}+B_{\epsilon}
\int_0^{\infty}\int_0^{\infty}du_1du_2 \;\; 2(F_b(u_1,u_2)-
F_{-b}(u_1,u_2))
\eea
thus the contribution of the subleading terms is twice that 
of solution 1. 

\subsection{Fake cusp calculation}

In this Appendix we evaluate the contribution of the fake cusps to the 6 point AdS amplitude
defined by $y_0=y_1|y_2|$.

But in order to do so 
we must select a method that will reproduce the correct behaviour for a correct cusp.

According to eq. 3.21 in \cite{buch} (see also equation 3.29 in \cite{am}), the contribution
to the string action from near a correct cusp is
\be
-iS_{i,i+1}(\epsilon)=\frac{\sqrt{\lambda_Dc_D}}{8\pi}\frac{\sqrt{1+\epsilon}}{(1+\epsilon/2)
^{1+\epsilon/2}2^{\epsilon/2}}(-s_{i,i+1})^{-\epsilon/2}\int_0^1\frac{dY_-dY_+}{(Y_-Y_+)
^{1+\epsilon/2}}
\ee
where $Y_{\pm}$ are coordinates parallel to the 2 momenta (sides of the cusps). The integration
in the original variables $y_{\pm}$ was from 0 to the values of the momentum, i.e. the length 
of the side of the cusp, except that the solution used was not the exact one for the 
polygon Wilson loop, but rather the approximate one for the infinite cusp. Note that the integral gives $(2/\epsilon)^2$ (it's the product of two identical
integrals). 

The integration above was done in $y_{\pm}=y_0\pm y_1$  variables (with $y_2$ added), since 
the solution used was an infinite cusp with lightlike boundary. But the fake 
cusp we are interested in has not only the 
lightlike boundary along $\tilde{y}_{\pm}=y_0\pm y_2$, but
also the boundary $y_0=0,y_2=0$, so clearly $\tilde{y}_{\pm}$ are not 
good integration variables for the cusp solution. Rather, we will use $y_1,y_2$. 

In order to understand the $y_1,y_2$ integration
procedure better, we will first analyze the b=0 solution, 
looking at both the usual (Alday-Maldacena) cusp and the new fake cusp. The solution is
\be
r^2=(1+\epsilon/2)(1-y_1^2)(1-y_2^2);\;\;\; y_0=y_1|y_2|
\ee
The square root in the action (\ref{action}) is (after a bit of algebra)
\be
\sqrt{1+\frac{\epsilon}{2}(y_1^2+y_2^2)}
\ee
Near a good cusp, e.g. $y_1=y_2=1$, we have 
\bea
&& r\simeq 2\sqrt{\delta y_1\delta y_2}\sqrt{1+\epsilon/2};\;\;\;
y_0\simeq 1-\delta y_1-\delta y_2\nonumber\\
&& {\cal L}=\frac{\sqrt{1+\epsilon}d\delta y_1 d\delta y_2}{(1+\epsilon /2)^{1+\epsilon/2}
(4\delta y_1\delta y_2)^{1+\epsilon/2}}
\eea
Then the action at the cusp is 
\be
-iS_{i,i+1}(\epsilon)=\frac{\sqrt{\lambda_Dc_D}}{8\pi}\frac{\sqrt{1+\epsilon}}{(1+\epsilon/2)
^{1+\epsilon/2}2^{\epsilon}}\int_0^1\frac{d\delta y_1d\delta y_2}{(\delta y_1\delta y_2)
^{1+\epsilon/2}}
\ee
Near a fake cusp, e.g. $y_2=0,y_1=1$, we have 
\bea
&&r\simeq\sqrt{2\delta y_1}\sqrt{1+\epsilon/2};\;\;\; y_0\simeq |\delta y_2|\nonumber\\
&& {\cal L}=\frac{\sqrt{1+\epsilon/2}d\delta y_1 d\delta y_2}{(1+\epsilon /2)^{1+\epsilon/2}
(2\delta y_1)^{1+\epsilon/2}}
\eea
and the action at the cusp is
\be
-iS_{i,i+1}(\epsilon)=\frac{\sqrt{\lambda_Dc_D}}{4\pi}\frac{\sqrt{1+\epsilon/2}}
{(1+\epsilon/2)
^{1+\epsilon/2}2^{\epsilon/2}}\int_0^1\frac{d\delta y_1}{(\delta y_1)^{1+\epsilon/2}}
\int_{-1}^1 d\delta y_2
\ee
which now contains a single divergent integral, so is of order $1/\epsilon$, not $1/\epsilon
^2$. 

Now we turn to the nonzero b case. For nonzero b, the equation of the 
Alday-Maldacena (4 point function) curve in $y_0,y_1,y_2,
r$ coordinates is obtained from (\ref{nonzerob}) by writing $\tanh u_1$, $\tanh u_2$ as a function of $y_1,y_2$ and substituting in $r, y_0$ with the result
\bea
&& y_0=\frac{\sqrt{1+b^2}}{2b}(1-\sqrt{1-4by_1y_2})\nonumber\\
&& r=\frac{1}{2b}\frac{\sqrt{[4b^2y_2^2-(1-\sqrt{1-4by_1y_2})^2][4b^2y_1^2-
(1-\sqrt{1-4by_1y_2})^2]}}{(1-\sqrt{1-4by_1y_2})}
\eea

For the 6 point function solution we replace everywhere $y_2$ by $|y_2|$. Near the fake 
corner $u_2=0, u_1=+\infty\leftrightarrow y_2=0, y_1=1$ we get (after some algebra)
\bea
&&y_0\simeq \sqrt{1+b^2}|\delta y_2|(1-\delta y_1+b|\delta y_2|)\nonumber\\
&&r\simeq \sqrt{1+\epsilon /2}\sqrt{2(\delta y_1-b|\delta y_2|)}(1-\frac{\delta y_1-b|\delta
y_2|}{4})
\eea
and, again after some algebra, we get the Lagrangian
\be
{\cal L}=\frac{\sqrt{1+\epsilon/2(1+b^2)}}{(1+\epsilon/2)^{1+\epsilon/2}2^{1+\epsilon/2}}
\frac{d\delta y_1d\delta y_2}{(d\delta y_1-b|\delta y_2|)^{1+\epsilon/2}}
\ee

Now we need to change to variables that are parallel to $k_4,k_5$. Since the momenta are
\be
k_4^{\mu}=\frac{1}{b+1}(b,1,...);\;\;\; k_5^{\mu}=
\frac{1}{b+1}(b,-1,...)
\ee
we get that the new variables $Y_1,Y_2$ that are parallel to $k_4,k_5$ and run from 0 to 1 
are defined as
\bea
&&\delta y_1=\frac{b}{b+1}(Y_1+Y_2);\;\;\; \delta y_2=\frac{1}{b+1}(Y_1-Y_2)
\Rightarrow \nonumber\\
&&d\delta y_1 d\delta y_2=\frac{2b}{(b+1)^2}dY_1dY_2;\;\;\; (\delta y_1-b|\delta y_2|)
=\frac{2b}{b+1}min\{ Y_1,Y_2\}
\eea

Then the action at the fake cusp is 
\bea
&&-iS_{i,i+1}=\frac{\sqrt{\lambda_Dc_D}}{2\pi}
\frac{\sqrt{1+\epsilon/2(1+b^2)}}{(1+\epsilon/2)^{1+\epsilon/2}
2^{1+\epsilon/2}}\int \frac{d\delta y_1 d\delta y_2}{(\delta y_1-b|\delta y_2|)^{1+\epsilon
/2}}\nonumber\\
&&=\frac{\sqrt{\lambda_Dc_D}}{2\pi}
\frac{\sqrt{1+\epsilon/2(1+b^2)}}{(1+\epsilon/2)^{1+\epsilon/2}
2^{1+\epsilon/2}}\frac{1}{b+1}(\frac{2b}{b+1})^{-\epsilon/2}\frac{-4}{\epsilon(1-\epsilon/2)}
\eea

If $\epsilon \ln b<1$ we obtain
\be
-iS_{i,i+1}\simeq
\frac{\sqrt{\lambda}}{2\pi}\frac{2}{\epsilon}\frac{1}{b+1}\left(\frac{\pi\mu}{a}\right)
^{\epsilon}(1+\frac{\epsilon}{4}(1+b^2)
+\frac{\epsilon}{2}(1-\ln \frac{b}{b+1}))
\ee

This contribution is indeed of order $1/\epsilon$ as we wanted (since we are missing the 
$1/\epsilon^2$ term), but the b dependence is incorrect.

\end{document}